\documentclass[cmp,final,numbook]{svjour}
\usepackage{graphics,amsbsy,times,amssymb}
\def\u{\stackrel{\mbox{\tiny $\rightarrow$}}{\boldsymbol u}}
\def\vv{\stackrel{\mbox{\tiny $\rightarrow$}}{\boldsymbol v}}
\def\dd{\delta}
\def\bphi{\bar{\phi}}
\def\bchi{\bar{\chi}}
\def\sphi{\phi^*}
\def\schi{\chi^*}
\def\rmg{{\rm g}} 
\def\Gl{{\cal G}}
\def\nb{\nabla}
\def\bnb{\bar{{\nabla}}}
\def\bR{\bar{R}}
\def\La{\Lambda}
\def\Up{\Upsilon}
\def\U{{\cal U}}
\def\bUp{\bar{\Up}}
\def\bLa{\bar{\La}}
\def\f{\varphi}
\def\r{\mathbb R}
\def\C{\boldsymbol{C}}
\def\G{\boldsymbol{\rmg}}
\def\pC{{}^{||}\C}
\def\oC{{}^{\perp}\C}
\def\bP{\boldsymbol{P}}
\def\bPi{\boldsymbol{\Pi}}
\def\bfeta{\boldsymbol{\eta}}
\def\bz{\bar{\boldsymbol{\theta}}}
\def\l{\lambda}
\def\z{\theta}

\def\fr{\frac}
\def\d{\partial}

\def\g{\gamma}
\def\Chr{\Gamma}

\def\bbg{\bar{\gamma}}
\def\vecX{{\mathfrak X}}
\def\xiv{\stackrel{\mbox{\tiny $\rightarrow$}}{\boldsymbol\xi}}
\def\lie{\pounds_{\xiv}}

\def\be{\begin{equation}}
\def\ee{\end{equation}}
\def\bnr{\begin{eqnarray*}}
\def\enr{\end{eqnarray*}}
\def\bea{\begin{eqnarray}}
\def\eea{\end{eqnarray}}

\begin{document}
\title{Bi-conformal vector fields and their applications to the characterization
of conformally separable pseudo-Riemannian manifolds}
\subtitle{New criteria for the existence of conformally flat foliations in pseudo-Riemannian manifolds}
\author{Alfonso Garc\'{\i}a-Parrado G\'omez-Lobo}                     
\institute{Departamento de F\'{\i}sica Te\'orica, 
Universidad del Pa\'{\i}s Vasco. Apartado 644, 48080 Bilbao 
(Spain)\\
\email{wtbgagoa@lg.ehu.es}}
\date{Received: date / Accepted: date}
% Add name of the expert who has communicated your paper

\titlerunning{Bi-conformal vector fields and their applications to the characterization of conformally ...}
\communicated{********}
\maketitle
\begin{abstract}
In this paper a thorough study of the normal form and the first integrability conditions arising from 
{\em bi-conformal vector fields} is presented. These new symmetry transformations were introduced in 
{\em Class. Quantum Grav.} \textbf{21}, 2153-2177
 and some of their basic properties were addressed there. Bi-conformal vector fields 
are defined on a pseudo-Riemannian manifold 
through the differential conditions $\lie P_{ab}=\phi P_{ab}$ and $\lie\Pi_{ab}=\chi\Pi_{ab}$ where 
$P_{ab}$ and $\Pi_{ab}$ are orthogonal and complementary projectors with respect to the 
metric tensor $\rmg_{ab}$. One of the main results of our study is 
the discovery of a new geometric characterization of {\em conformally separable spaces with conformally flat leaf metrics}
similar to the vanishing of the Weyl tensor for conformally flat metrics. 
This geometric characterization seems to carry over to any pseudo-Riemannian manifold admitting conformally 
flat foliations which would open the door to 
the systematic searching of these type of foliations in a given pseudo-Riemannian metric. 
Other relevant aspects such as the existence of 
invariant tensors under the finite groups generated by these transformations are also addressed.  
\end{abstract}
\section{Introduction}
\label{intro}
The research of symmetry transformations in Differential Geometry and General Re\-la\-ti\-vity has been an 
important subject during 
the years.
Here by symmetries we mean a group of transformations of a given pseudo-Riemannian manifold complying with certain geometric 
property. By far the most studied symmetries are isometries and conformal transformations which are defined 
through the conditions
\be
\lie\rmg_{ab}=0,\ \ \lie\rmg_{ab}=2\phi\rmg_{ab},
\label{n-zero}
\ee
where $\rmg_{ab}$ is the metric tensor of the manifold, $\xiv$ is the {\em infinitesimal generator} of the transformation and
$\phi$ is a function which we will call {\em gauge} of the symmetry (this terminology was first 
employed in \cite{SERGI} and it will be explained later). 
Infinitesimal generators of these symmetries are known as Killing vectors and conformal Killing vectors. 
As is very simple to check they are a Lie algebra with respect to the Lie bracket of vector fields 
and the transformations generated by these vector fields give rise to subgroups of the diffeomorphism group.  

Important questions are the possible dimensions of these Lie algebras and the geometric characterizations of spaces admitting the
symmetry. The general answer to these questions can in principle be obtained by solving the differential conditions 
written above although for general enough cases the explicit evaluation of such solutions gets too complex and other methods are 
required. Notwithstanding these difficulties we can obtain easily from the differential conditions the cases in which the Lie 
algebras are finite dimensional, the greatest dimension of these finite Lie algebras and geometric characterizations of the 
spaces admitting these Lie algebras as solutions. This is done by finding the {\em normal form} of the above equations 
(if such form exists) and the complete integrability conditions coming from this set of equations.
In this way we deduce that isometries are always finite dimensional whereas 
conformal motions are finite dimensional iff the space dimension is greater or equal than three. 
The spaces in which the greatest 
dimension is achieved are constant curvature and conformally flat spaces respectively
and as is very well known they are characterized by the geometric conditions
$$
R^a_{\ bcd}=\fr{R}{n(n-1)}(\dd^a_{\ c}\rmg_{bd}-\dd^a_{\ d}\rmg_{bc}),\ \ C^a_{\ bcd}=0,\ n>3,
$$ 
where $n$ is the dimension of the manifold, $R^a_{\ bcd}$ is the curvature tensor, $R$ the scalar curvature and $C^a_{\ bcd}$ 
the Weyl tensor\footnote{In the case of dimension three the Weyl tensor is replaced 
by a three rank tensor called the Cotton-York tensor.}

The procedure followed for isometries and conformal motions is carried over to other symmetries such as 
linear and affine collineations and conformal collineations (see \cite{YANO,LIBROHALL} for a very good 
account of this). However, little research has been done for symmetries different from these mostly because the cases under 
consideration were infinite dimensional {\em generically}. This means that it is not possible to obtain a normal set of 
equations out of the differential conditions (see section \ref{normal-dimension}) which greatly complicates matters.

In cite \cite{BI-CONFORMAL} we put forward a new symmetry transformation for general pseudo-Riemannian manifolds. 
Infinitesimal generators of these symmetries (bi-conformal vector fields) fulfill the differential conditions 
\be
\lie P_{ab}=\phi P_{ab},\ \ \lie\Pi_{ab}=\chi\Pi_{ab},
\label{n-uno}
\ee
where $P_{ab}$ and $\Pi_{ab}$ are orthogonal and complementary projectors with respect to the metric tensor $\rmg_{ab}$ and 
$\phi$, $\chi$ are the gauges of the symmetry. These are functions which, as happened in the conformal case, depend on the 
vector field $\xiv$ so a solution of (\ref{n-uno}) is formed by $\xiv$ itself and the gauges $\phi$ and $\chi$ (we will 
usually omit the dependence on $\xiv$ in the gauges).
The finite transformations generated by bi-conformal vector fields are called bi-conformal transformations. 
In a sense, these symmetries can be regarded as conformal 
transformations with respect to both $P_{ab}$ and $\Pi_{ab}$ so we can expect that some properties of these will resemble 
those of conformal transformations. In \cite{BI-CONFORMAL} it was shown that bi-conformal vector fields comprise a Lie algebra 
under the Lie bracket and that this algebra is finite dimensional
if none of the projectors has algebraic rank one or two being the greatest dimension
$$
N=\fr{1}{2}p(p+1)+\fr{1}{2}(n-p)(n-p+1),
$$
with $p$ the algebraic rank of one of the projectors. We provided also explicit examples in which this dimension is achieved,
namely bi-conformally flat spaces which in local coordinates $x=\{x^1,\dots,x^n\}$ look like 
$(\alpha,\beta=1,\dots p,\ A,B=p+1,\dots n)$ 
$$
ds^2=\Xi_1(x)\eta_{\alpha\beta}dx^{\alpha}dx^{\beta}+\Xi_2(x)\eta_{AB}dx^Adx^B,\ \Xi_1,\ \Xi_2\in C^3,
$$
where $\eta_{\alpha\beta}$, $\eta_{AB}$ are flat metrics depending only on the coordinates $x^{\alpha}$ and $x^A$ respectively.
These spaces play the same role for bi-conformal vector fields as conformally flat spaces or spaces of 
constant curvature do for the classical symmetries.
An open question was the finding of a geometric characterization for bi-conformally flat spaces similar to those 
of the spaces of constant curvature or conformally flat spaces. In the scheme developed in \cite{BI-CONFORMAL} this
sort of characterization could not be extracted due to the complexity of the calculations and it had to be postponed. 
This matter and other important properties of bi-conformal vector fields are dealt with in the present work.

In this paper we perform the full calculation of the normal form and the first integrability conditions 
for the equations (\ref{n-uno}). In our previous work we were able to get the normal form for these 
equations but it turned out to be rather messy and relevant geometric information could not be obtained. 
This was so because all these calculations were done using the covariant derivatives arising from the metric
connection which is not adapted to the calculations. Here we show that the definition of a new symmetric connection 
greatly simplifies the calculations making it possible to get a simpler form for the normal system and 
to work out its integrability conditions. In particular a geometric characterization for bi-conformally flat spaces 
in terms of certain tensors $T^a_{\ bcd}$ and $T_{abc}$ constructed from the projectors $P_{ab}$ and $\Pi_{ab}$ is derived.
We prove the remarkable result that these both tensors are zero 
if and only if the space is bi-conformally flat being $P_{ab}$ and $\Pi_{ab}$ the conformally flat pieces 
in which the metric is split up (here neither of the projectors can have algebraic rank three). Along the way we obtain 
other interesting results such as the geometric characterization of {\em conformally separable} pseudo-Riemannian manifolds 
(in this case this comes through the vanishing of $T_{abc}$)
and the existence of {\em geometric invariants} similar to the Riemann tensor or the Weyl tensor for isometries or 
conformal vector fields. We can also give geometric conditions under which a conformally separable pseudo-Riemannian 
manifold admits conformally 
flat leaf metrics (see definitions \ref{metric-separable} and \ref{classification} for explanations about this 
terminology). These conditions take the form 
$$
\ P^a_{\ r}P_b^{\ s}P_c^{\ q}P_d^{\ t}T^r_{\ sqt}=0.
$$
In fact these conditions appear to be true in the case of general pseudo-Riemannian manifolds admitting conformally 
flat slices as we show by means of explicit examples. This is very interesting because it would enable us to systematically
search foliations by conformally flat hypersurfaces in general pseudo-Riemannian manifolds (and even decide whether 
these foliations exist or not).
   
The outline of the paper is as follows: section \ref{definitions} introduces the basic notation and definitions. 
 In section \ref{connection} we define a new symmetric connection (bi-conformal connection) and set its main properties.
Section \ref{normal-dimension} presents the calculation of the normal form associated to (\ref{n-uno}) and the calculation
of the maximum dimension of any finite dimensional Lie algebra of bi-conformal vector fields is carried out. The first 
integrability conditions of the equations comprising the normal form are worked out in section \ref{first-integrability} and
in these calculations a family of invariant tensors under bi-conformal transformations comes up naturally. The complete 
integrability conditions are the subject of section \ref{complete} and they lead to the geometric characterizations 
of conformally separable and bi-conformally flat pseudo-Riemannian manifolds explained above. These characterizations 
are spelt out in detail in section \ref{sufficient} where we provide a full account of all the possible cases and 
we point out that some of them correspond to the complete integrability conditions. Finally in section \ref{examples} 
we show explicit examples of how to use the geometric characterizations found in the paper and we hint how these conditions 
may be extended to metrics which are not conformally separable.   

\section{Bi-conformal vector fields and bi-conformal transformations}
\label{definitions}
Let us start by setting our notation and conventions for the paper 
(they are though rather general and obvious so the paper should present no difficulties in 
this point). We work on a differentiable manifold $V$ in which a $C^{\infty}$
metric $\rmg_{ab}$ of arbitrary signature has been defined (pseudo-Riemannian manifold). 
Vectors and vector fields are denoted with arrowed characters 
$\u$, $\vv$
 (we leave to the context the distinction between each of these 
entities) when expressed in coordinate-free notation whereas 
1-forms are written in bold
characters ${\boldsymbol u}$. Sometimes this same notation will be employed for other higher rank objects
such as contravariant and covariant tensors. 
 Indexes of tensors are represented by lowercase Latin characters $a$, $b,\dots $ 
and the metric $\rmg_{ab}$ or its inverse $\rmg^{ab}$ are used to respectively raise 
or lower indexes. Rounded and square brackets are used for symmetrization and antisymmetrization respectively
and whenever a group of indexes is enclosed between strokes they are excluded from the symmetrization 
or antisymmetrization operation.
 Partial derivatives with respect to local coordinates are $\d_a\equiv\d/\d x^a$.
The metric connection associated to $\rmg_{ab}$ is $\g^{a}_{\ bc}$ (Ricci rotation coefficients) reserving
the symbol $\Gamma^a_{\ bc}$ only for the Christoffel symbols, namely, the connection components
in a natural basis. The covariant derivative and the Riemann tensor constructed from 
this connection are denoted by $\nb$ and $R^a_{\ bcd}$ respectively being our convention for the 
Riemann tensor
$$
R^a_{\ bcd}\equiv\d_c\Chr^a_{\ db}-\d_d\Chr^a_{\ cb}+\Chr^a_{\ rc}\Chr^r_{\ db}-\Chr^a_{\ rd}\Chr^r_{\ cb}.
$$
Under this convention Ricci identity becomes 
$$
\nb_b\nb_cu^a-\nb_c\nb_bu^a=R^a_{\ rbc}u^r,\ \ \nb_b\nb_cu_a-\nb_c\nb_bu_a=-R^r_{\ abc}u_r.
$$ 
All the above relations are still valid for a non-metric symmetric connection.
The push-forward and pull-back of elements of the tensor bundle $T^r_s(V)$ (bundle of 
$r$-con\-tra\-va\-ri\-ant $s$-covariant tensors) or its sections
under a diffeomorphism $\Phi:V\rightarrow V$ are represented by the standard notation 
$\Phi_{*}\!\!\!\u$ and $\Phi^*{\boldsymbol u}$ respectively. We shall use $\ ^{*}\Phi^*{\boldsymbol T}$ if 
${\boldsymbol T}$ is a tensor with contravariant and covariant indexes. 
 
The set of smooth vector fields of the manifold
$V$ is denoted by $\vecX(V)$. This is an infinite dimensional Lie algebra  which is sometimes regarded as the Lie
algebra of the group of diffeomorphisms of the manifold $V$. Finally the Lie derivative operator 
with respect to a vector field $\xiv$ is $\lie$.

One of the main subjects of this paper is the study of bi-conformal vector fields whose definition 
given in \cite{BI-CONFORMAL} we recall here.
\begin{definition}
A $C^2$ vector field $\xiv$ on $V$ is said to be a {\em bi-conformal vector field} 
if it fulfills the condition
\be 
\lie P_{ab}=\phi P_{ab},\ \ \ 
\lie\Pi_{ab}=\chi\Pi_{ab},
\label{bi-conformal}
\ee
for some functions $\phi$, $\chi\in C^2(V)$. 
\label{Bi-conformal}
\end{definition}
$P_{ab}$ and $\Pi_{ab}$ are elements of the tensor
bundle $T^0_2(V)$ such that at each point $x\in V$ they form a pair of orthogonal and 
complementary projectors with respect to the metric tensor $\rmg_{ab}|_x$ which leads to 
$P_{ab}+\Pi_{ab}=\rmg_{ab}$, $P_{ap}P^{p}_{\ b}=P_{ab}$, $\Pi_{ap}\Pi^{p}_{\ b}=\Pi_{ab}$ and 
$P_{ap}\Pi^p_{\ b}=0$. 
Equation (\ref{bi-conformal}) can be re-written in a 
number of equivalent ways if we use the square root of the metric tensor $S_{ab}$ defined in terms of the 
projectors $P_{ab}$ and $\Pi_{ab}$ \cite{BI-CONFORMAL}
\be
S_{ab}\equiv P_{ab}-\Pi_{ab}, \Longleftrightarrow P_{ab}=\fr{1}{2}(\rmg_{ab}+S_{ab}),\ \ 
\Pi_{ab}=\fr{1}{2}(\rmg_{ab}-S_{ab}),\Longrightarrow S_{ap}S^p_{\ b}=\rmg_{ab}, 
\label{square-root}
\ee
 getting
\be
\lie\rmg_{ab}=\alpha\rmg_{ab}+\beta S_{ab},\ \ 
\lie S_{ab}=\alpha S_{ab}+\beta\rmg_{ab},\ \ 
\alpha=\fr{1}{2}(\phi+\chi),\ \ \beta=\fr{1}{2}(\phi-\chi).
\label{second}
\ee
From equation (\ref{square-root}) 
we deduce that both projectors are fixed by the square root $S_{ab}$
so we may use the latter instead of the projectors when working with a given set of 
bi-conformal vector fields. 
Following \cite{BI-CONFORMAL} the set of bi-conformal vector fields possessing $S_{ab}$ as the 
associated square root will be denoted by $\Gl(S)$. 
In this paper only expressions involving $P_{ab}$ and $\Pi_{ab}$ will be used in our calculations.

The next set of relations come straight away from (\ref{bi-conformal})
\be
\lie P^a_{\ b}=\lie \Pi^a_{\ b}=0,\ \ \lie P^{ab}=-\phi P^{ab},\ \lie\Pi^{ab}=-\chi\Pi^{ab}.
\ee

A very important property of $\Gl(S)$ 
is that it forms a Lie subalgebra of $\vecX(V)$
 (proposition 5.2 of \cite{BI-CONFORMAL}) which can be finite or infinite dimensional.
Conditions upon the tensor $S_{ab}$ (or equivalently the projectors) for this Lie algebra to be finite dimensional 
were given in \cite{BI-CONFORMAL} and they will be re-derived in section \ref{normal-dimension} in a more efficient 
way. Observe that the functions $\alpha$ and 
$\beta$ appearing in definition \ref{Bi-conformal} (or $\phi$ and $\chi$)
do depend on the bi-conformal vector field (this dependence can be
dropped if we work with a single bi-conformal vector field but it should be added when working with Lie algebras of 
bi-conformal vector fields). In the latter case $\alpha$ and $\beta$ are called {\em gauge functions}
(see \cite{SERGI} for an explanation of this terminology).   

The finite counterpart of bi-conformal vector fields ({\em bi-conformal transformations}) was
also presented in section 4 of \cite{BI-CONFORMAL} but for the sake of completeness we repeat again
the definitions here referring the interested reader to that paper for further details.  
\begin{definition}
Under the assumptions of definition \ref{Bi-conformal}
a diffeomorphism $\Phi:V\rightarrow V$ is called a bi-conformal transformation if 
the conditions $\Phi^*P_{ab}=\alpha_1P_{ab}$, $\Phi^*\Pi_{ab}=\alpha_2\Pi_{ab}$ are satisfied 
for some functions $\alpha_1$, $\alpha_2\in C^0(V)$.
\label{finite-form}
\end{definition}
In this paper the majority of our calculations
will only involve bi-conformal vector fields and we will draw our results from infinitesimal 
equations instead of expressions arising from bi-conformal relations. 

\section{The bi-conformal connection}
\label{connection}
As we have commented in the introduction
the ordinary metric connection $\gamma^a_{\ bc}$ defined in the 
standard way from the metric is not suitable to study the normal form and the integrability 
conditions coming from the differential condition (\ref{bi-conformal}) as they result in 
rather cumbersome expressions. In order to proceed further in our study we are going to 
show next that the definition of a new symmetric connection greatly simplifies the normal form 
calculated in \cite{BI-CONFORMAL} and what is more, it allows us to work out thoroughly the 
complete integrability conditions arising from this normal form.

To start with we recall some identities satisfied by any bi-conformal vector field $\xiv$
which were obtained in \cite{BI-CONFORMAL}. These identities are in fact linear combinations of 
the first covariant
derivative of (\ref{bi-conformal}) and we also indicate briefly 
how they are obtained as this information
will be needed later. Using equation (\ref{lie-connection}) we easily obtain the Lie derivative 
of the metric connection $\g^a_{\ bc}$ ($\phi_b\equiv\d_b\phi$, $\chi_b\equiv\d_b\chi$)
\be
\lie\g^a_{bc}=\fr{1}{2}(\phi_bP^a_{\ c}+\phi_cP^a_{\ b}-\phi^aP_{bc}+
\chi_b\Pi^a_{\ c}+
\chi_c\Pi^a_{\ b}-\chi^a\Pi_{cb}+(\phi-\chi)M^a_{\ bc}),
\label{3}
\ee
where the tensor $M_{abc}$ is defined by
\be
M_{abc}\equiv\nb_bP_{ac}+\nb_cP_{ab}-\nb_aP_{bc}.
\ee
 
The Lie derivative of $M_{abc}$ can be worked out by means of (\ref{lie-conmmutation}) 
getting
\bea
\lie M_{abc}=\phi M_{abc}+(\chi-\phi)P_{ap}M^{p}_{\ bc}-       
P_{bc}\Pi_{ap}\phi^p+\Pi_{cb}P_{ap}\chi^p=\nonumber\\
=\chi M_{abc}+(\phi-\chi)\Pi_{ap}M^p_{\ bc}-       
P_{bc}\Pi_{ap}\phi^p+\Pi_{cb}P_{ap}\chi^p\label{9},
\eea
from which, projecting down with $P^{bc}$ and $\Pi^{bc}$, we deduce 
\be
\lie E_a=-p\Pi_{ap}\phi^p,\ \ \ \lie W_a=(p-n)P_{ap}\chi^p,
\label{18}
\ee
with the definitions 
$$
E_a\equiv M_{acb}P^{cb},\ \ W_a\equiv-M_{acb}\Pi^{cb},\ p=P^a_{\ a}.
$$
The following algebraic properties of the tensors $E_a$ and $W_a$ are useful
\be
\Pi_{ac}E^c=E_a,\  P_{ac}W^c=W_a,\ 
0=P^{ab}E_b=\Pi^{ab}W_b.
\label{vwe}
\ee
Now, we plug (\ref{18}) into (\ref{9}) yielding
\bea
\lie\left(M_{acb}-\fr{1}{p}E_aP_{bc}+\fr{1}{n-p}W_a\Pi_{cb}\right)=
\phi\left(\Pi_{ap}M^p_{\ cb}-\fr{E_aP_{bc}}{p}\right)+\nonumber\\
\hspace{15mm}+\chi\left(P_{ap}M^p_{\ bc}+\fr{\Pi_{cb}W_a}{n-p}\right).
\label{integral-constraint}
\eea 
This equation can be written in a more compact form
\be
\lie T_{abc}=(\phi\Pi_{ar}+\chi P_{ar})T^r_{\ bc}=\phi B_{abc}+\chi A_{abc},
\label{compact}
\ee
where the definitions of the tensors $T_{abc}$, $A_{abc}$, $B_{abc}$ are
\bea
T_{abc}\equiv M_{abc}+\fr{1}{n-p}W_a\Pi_{bc}-\fr{1}{p}E_aP_{bc},\label{tensor-t}\\ 
A_{abc}\equiv P_{a}^{\ d}T_{dbc}=P_a^{\ d}M_{dcb}+\fr{1}{n-p}W_a\Pi_{cb},\label{splitting-1}\\
B_{abc}\equiv\Pi_{a}^{\ d}T_{dbc}=\Pi_{a}^{\ d}M_{dcb}-\fr{1}{p}E_aP_{cb}.\label{splitting-2}
\eea
Using equation (\ref{compact}) we can calculate the Lie derivatives of $A^a_{\ bc}$ and 
$B^a_{\ bc}$
\be
\lie A^a_{\ bc}=(\chi-\phi)A^a_{\ bc},\ \ \lie B^a_{\ bc}=(\phi-\chi)B^a_{\ bc},
\label{AyB}
\ee
 a relation which shall be used later. See \cite{BI-CONFORMAL} for further properties of these 
tensors.

Let us now use all this information to write the Lie derivative of the connection in a convenient 
way. Note that in equation (\ref{9}) $\phi_a$ and $\chi_a$ appear projected with $P^a_{\ b}$ and 
$\Pi^a_{\ b}$ respectively suggesting that it could be interesting to write any derivative of 
$\phi$ and $\chi$ decomposed in longitudinal and transverse parts 
\be
\sphi_a\equiv \Pi_{ab}\phi^b,\ \bphi_a\equiv P_{ab}\phi^b,\ \schi_a\equiv P_{ab}\chi^b,\ \ 
\bchi_a\equiv\Pi_{ab}\chi^b.
\ee
If we perform this decomposition in equation (\ref{3}) and replace the longitudinal terms 
$\sphi_a$ and $\schi_a$ by means of (\ref{18}) we get the relation
\bea
& &\hspace{-.5cm}\lie\left(\g^a_{\ bc}+\fr{1}{2p}(E_bP^a_{\ c}+E_cP^a_{\ b}-P_{bc}E^a)+
\fr{1}{2(n-p)}(W_b\Pi^a_{\ c}+W_c\Pi^a_{\ b}-W^a\Pi_{cb})\right)=\nonumber\\
&=&\fr{1}{2}(\bphi_bP^a_{\ c}+\bphi_cP^a_{\ b}-\bphi^aP_{bc}+\bchi_b\Pi^a_{\ c}+\bchi_c\Pi^a_{\ b}-
\bchi^a\Pi_{cb})+\fr{1}{2}(\phi-\chi)T^a_{\ bc},
\label{connection1}
\eea
but from (\ref{AyB}), (\ref{splitting-1}) and (\ref{splitting-2}) we easily deduce  
$$
\lie(A^a_{\ bc}-B^a_{\ bc})=(\chi-\phi)T^a_{\ bc},
$$
hence equation (\ref{connection1}) becomes, after some simplifications
\bea
& &\hspace{-.7cm}
2\lie\left(\g^a_{\ bc}+\fr{1}{2p}(E_bP^a_{\ c}+E_cP^a_{\ b})+\fr{1}{2(n-p)}(W_b\Pi^a_{\ c}+W_c\Pi^a_{\ b})+
\fr{1}{2}(P^a_{\ p}-\Pi^a_{\ p})M^p_{\ cb}\right)=\nonumber\\
&=&\bphi_bP^a_{\ c}+\bphi_cP^a_{\ b}-\bphi^aP_{cb}+\bchi_b\Pi^a_{\ c}+\bchi_c\Pi^a_{\ b}-\bchi^a\Pi_{cb}.
\label{der-connection}
\eea
The geometric object inside the Lie derivative, denoted by $\bbg^a_{\ bc}$,
 is the sum of the metric connection $\g^a_{\ bc}$ plus the rank-3 tensor 
\be
L^a_{\ bc}\equiv\fr{1}{2p}(E_bP^a_{\ c}+E_cP^a_{\ b})+\fr{1}{2(n-p)}(W_b\Pi^a_{\ c}+W_c\Pi^a_{\ b})+
\fr{1}{2}(P^a_{\ p}-\Pi^a_{\ p})M^p_{\ bc},
\label{labc}
\ee
so it is clear that it represents a new linear connection. 
As we will see during our calculations this linear connection is fully adapted to the calculations involving 
bi-conformal vector fields and it will be extensively used in this paper. 

\begin{definition}[bi-conformal connection]
The connection whose components are gi\-ven by $\bbg^a_{\ bc}$ is called bi-conformal connection. 
The covariant derivative constructed from the bi-conformal connection shall be denoted by $\bnb$ and the 
curvature tensor constructed from it by $\bar{R}^a_{\ bcd}$.
\label{biconformal}
\end{definition}

Since $L^a_{\ bc}$ is symmetric in the indexes $bc$, we see that the bi-conformal connection is symmetric 
so it has no torsion and all
the identities involving only the covariant derivative $\bnb$ or the curvature $\bR^a_{\ bcd}$ remain the same
as for the case of a metric connection. However, this connection does not in general stem from a metric tensor 
as we will see later in explicit examples.
This means that certain symmetries of the curvature tensor of a metric 
connection are absent for $\bR^a_{\ bcd}$. 
We recall that for a symmetric connection the Riemann tensor is only antisymmetric in the last pair of indexes.
Bianchi identities however, remain the same as for the case of a 
metric connection.

It is not very difficult to derive an identity relating the curvature tensor calculated from the 
bi-con\-for\-mal connection and the 
curvature tensor associated to the connection $\g^a_{\ bc}$
\be
\bR^a_{\ bcd}=R^a_{\ bcd}+2\nb_{[c}L^a_{\ d]b}+2L^a_{\ r[c}L^r_{\ d]b},
\label{curvatures-relation}
\ee
being this a thoroughly general identity for two symmetric connections 
$\bbg^a_{\ bc}$ and $\g^a_{\ bc}$ differing in a tensor $L^a_{\ bc}$ (\cite{SCHOUTEN}, p. 141).

The relation between the covariant derivatives $\bnb$ and $\nb$ acting on any tensor 
$X^{a_1\dots a_p}_{\ b_1\dots b_q}$ is 
\bea
\bnb_aX^{a_1\dots a_r}_{\ b_1\dots b_q}=\nb_aX^{a_1\dots a_r}_{\ b_1\dots b_q}+\sum_{s=1}^rL^{a_s}_{\ ac}
X^{a_1\dots a_{s-1}ca_{s+1}\dots a_r}_{\ b_1\dots b_q}-\nonumber\\
-\sum_{s=1}^qL^{c}_{ab_s}X^{a_1\dots a_r}_{\ b_1\dots b_{s-1}cb_{s+1}\dots b_q},
\label{barcon-bar}
\eea
which again has general validity for two symmetric connections whose difference is a tensor $L^a_{\ bc}$ 
\cite{EISENHART}.
As a first application of this identity we may compare the covariant derivatives of the tensor $L^a_{\ bc}$ 
which leads us to the identity
$$
\bnb_{[a}L^b_{\ c]d}=\nb_{[a}L^b_{\ c]d}+2L^b_{\ r[a}L^r_{\ c]d},
$$ 
from which we can rewrite (\ref{curvatures-relation}) in terms of $\bnb$  
\be
\bR^a_{\ bcd}=R^a_{\ bcd}-2\bnb_{[c}L^a_{\ d]b}+2L^a_{\ r[c}L^r_{\ d]b}.
\label{curvature-connection}
\ee
Of course this last equation could have been obtained
 from (\ref{curvatures-relation}) by means of the replacements
$L^a_{\ bc}\rightarrow -L^a_{\ bc}$ and $\nb_a\rightarrow-\bnb_a$.
\begin{example}
To realize the importance of bi-conformal connection in future calculations, let us calculate 
its components for a {\em conformally separable} pseudo-Riemannian manifold 
(see definition \ref{classification}) given in local coordinates $x\equiv\{x^1,\dots,x^n\}$ by
\be
ds^2=\Xi_1(x)G_{\alpha\beta}(x^{\delta})dx^{\alpha}dx^{\beta}+\Xi_2(x)G_{AB}(x^C)dx^Adx^B.
\ee
Here Greek indexes range from 1 to $p$ and uppercase Latin indexes from $p+1$ to $n$ so the 
metric tensors $G_{\alpha\beta}$ and $G_{AB}$ are of rank $p$ and $n-p$ respectively.
The non-zero Christoffel symbols for this metric are
\bnr
\Chr^{\alpha}_{\ \beta\g}=\fr{1}{2\Xi_1}G^{\alpha\rho}(\d_{\beta}(\Xi_1G_{\alpha\rho})+
\d_{\g}(\Xi_1G_{\rho\beta})-\d_{\rho}(\Xi_1G_{\beta\g})),\\
\Chr^{A}_{\ BC}=\fr{1}{2\Xi_1}G^{AD}(\d_{B}(\Xi_1G_{CD})+
\d_{C}(\Xi_1G_{DB})-\d_{D}(\Xi_1G_{BC})),\\
\Chr^{\alpha}_{\beta A}=\fr{1}{2\Xi_1}\delta^{\alpha}_{\ \beta}\d_A\Xi_1,\ \ 
\Chr^A_{\ B\alpha}=\fr{1}{2\Xi_2}\delta^A_{\ B}\d_{\alpha}\Xi_2.
\enr
from which the only nonvanishing components of $M_{abc}$, $E_a$, $W_a$ are
\bea
M_{\alpha AB}=\d_{\alpha}(\Xi_2G_{AB}),\ \ 
M_{A\alpha\beta}=-\d_{A}(\Xi_1G_{\alpha\beta}),\nonumber\\
E_{A}=-\d_{A}\log|\mbox{det}(\Xi_1^pG_{\alpha\beta})|,\ \ 
W_{\alpha}=-\d_{\alpha}\log|\mbox{det}(\Xi_2^{n-p}G_{AB})|.
\eea
Therefore we get for the components of the bi-conformal connection
\bea
\bar{\Chr}^{\alpha}_{\ \beta\phi}&=&\fr{1}{2\Xi_1}(\delta^{\alpha}_{\ \beta}\d_{\phi}\Xi_1+\delta^{\alpha}_{\ \phi}\d_{\beta}\Xi_1-
G^{\alpha\rho}G_{\beta\phi}\d_{\rho}\Xi_1)+\Chr^{\alpha}_{\ \beta\phi}(G),\nonumber\\
\bar{\Chr}^A_{\ BC}&=&\fr{1}{2\Xi_2}(\delta^{A}_{\ B}\d_{C}\Xi_2+\delta^{A}_{\ C}\d_{B}\Xi_2-
G^{AR}G_{BC}\d_{R}\Xi_2)+\Chr^{A}_{\ BC}(G)\nonumber\\
\bar{\Chr}^{\alpha}_{\ \beta C}&=&\bar{\Chr}^{A}_{\ B\phi}=0,
\label{new-components}
\eea
where $\Chr^{\alpha}_{\ \beta\phi}(G)$ and $\Chr^{A}_{\ BC}(G)$ are the Christoffel symbols of the metrics 
$G_{\alpha\beta}$ and $G_{AB}$ respectively. From the above formulae we deduce that the bi-conformal connection is 
fully adapted to conformally separable pseudo-Riemannian manifolds because its components clearly split in two 
parts being each of them the Christoffel symbols of the metrics $G_{\alpha\beta}$, $G_{AB}$ plus terms 
involving the derivatives of the factors $\Xi_1$ and $\Xi_2$. We will take advantage of this property in section \ref{sufficient}
where we will provide invariant characterizations of conformally separable and {\em bi-conformally flat}
pseudo-Riemannian manifolds. 
\label{example1}
\end{example}

We calculate next the covariant derivative with respect to the bi-conformal connection of a number of 
tensors.

\begin{proposition}
The following identities hold true 
\bea
\bnb_aP_{bc}&=&\nb_aP_{bc}-\fr{1}{p}E_aP_{bc}-\fr{1}{2p}(E_bP_{ac}+E_cP_{ab}) 
-\fr{1}{2}(P_{cp}M^p_{\ ab}+P_{bp}M^p_{\ ac}),\label{identity-1}\\
2\bnb_aP^b_{\ c}&=&2\nb_aP^b_{\ c}+P^{bq}P^r_{\ c}M_{qra}-\Pi^{bq}P^r_{\ c}M_{qra}-
P^b_{\ q}M^q_{\ ac}+\fr{1}{n-p}W_c\Pi^b_{\ a}-
\fr{1}{p}E_cP^b_{\ a},\nonumber\\
\label{identity-2}\\
\bnb_aP^{bc}&=&\nb_aP^{bc}+\fr{1}{p}E_aP^{bc}+\fr{1}{2(n-p)}(W^c\Pi^b_{\ a}+W^b\Pi^c_{\ a})-
\fr{1}{2}(M^b_{\ ar}P^{rc}+M^c_{\ ar}P^{rb}),\nonumber\\
\label{identity-3}
\eea
and all the identities formed with the replacements $P_{ab}\rightarrow\Pi_{ab}$, $p\rightarrow n-p$.
\label{identity-set}
\end{proposition}
\begin{proof}
All these identities are proven by means of (\ref{barcon-bar}) and the use of  
 properties (\ref{vwe}).\qed 
\end{proof}
Using the above properties we can get more interesting identities to be used later on.
\bea
\bnb_aP^{ab}=\bnb_a\Pi^{ab}=\bnb_aP^a_{\ b}=\bnb_a\Pi^a_{\ b}=0,\label{contraction}\\
P^{bc}\bnb_aP_{bc}=-E_a,\ \ P_{bc}\bnb_aP^{bc}=E_a,\label{EW}\\
\Pi^{bc}\bnb_a\Pi_{bc}=-W_a,\ \ \Pi_{bc}\bnb_a\Pi^{bc}=W_a,\label{WE}\\
P^d_{\ r}\bnb_b\Pi^r_{\ d}=\Pi^d_{\ r}\bnb_bP^r_{\ d}=P^{dr}\bnb_{d}P_{rb}=\Pi^{dr}\bnb_d\Pi_{rb}=0.\label{PPi}
\eea
Note that index raising and lowering do not commute with $\bnb$ so we must be very careful when we 
raise or lower indexes in tensor expressions involving $\bnb$.

\section{Normal form and dimension of maximal Lie algebras of bi-conformal fields}
\label{normal-dimension}
We turn now our attention to the calculation of the full normal form coming from 
the differential conditions (\ref{Bi-conformal}). A detailed explanation of the general procedure and 
relevance of this calculation for a general symmetry can be found in \cite{EISENHARTII,BI-CONFORMAL} 
(see also \cite{YANO} 
for the calculation in the cases of the most studied symmetries in General Relativity such as isometries 
and conformal motions). Before starting the 
calculation and for the sake of completeness let us give a very brief sketch of the whole procedure. 
We must differentiate the condition (\ref{Bi-conformal}) a number of times in such a way that we get enough 
equations to isolate the derivatives of certain variables (system variables) in terms of themselves (this is 
achieved typically by means of the resolution of a linear system of equations). 
The so obtained derivatives give rise to the normal form associated to our symmetry. 
Along the differentiating process one
may obtain equations whose linear combinations no longer contain derivatives of the system variables 
(constraints). Examples of such constraints in our case are the differential conditions themselves and 
(\ref{18}) (actually these are the only constraints as we will show in $\S$\ref{constraints}). 
   
As we will see it is possible to meet cases in which the normal form cannot be achieved. This means that 
one cannot obtain enough equations to isolate all the derivatives obtained through the derivation process.
The main implication of this is that the Lie algebra of vector fields fulfilling the starting differential 
condition is infinite dimensional as opposed to the case in which there is such a normal form. Therefore the 
calculation of the normal form allows us to tell apart the cases with an infinite dimensional Lie algebra of 
vector fields from the solutions representing finite dimensional Lie algebras. In the latter case we can even go
further and determine the highest dimension of these Lie algebras as the total number of system variables 
minus the number of linearly independent constraints.  

We start out our calculation with the substitution of (\ref{der-connection}) into (\ref{lie-xi}) which yields
\be
\bnb_b\Psi_c^{\ a}+\xi^d\bar{R}^a_{\ cdb}=\fr{1}{2}(\bphi_bP^a_{\ c}+\bphi_cP^a_{\ b}-\bphi^aP_{cb}+
\bchi_b\Pi^a_{\ c}+\bchi_c\Pi^a_{\ b}-\bchi^a\Pi_{cb}),\ \Psi_c^{\ a}\equiv\bnb_c\xi^a.
\label{psi}
\ee
Next we replace in (\ref{lie-curvature}) the Lie derivatives of the bi-conformal connection by their 
expressions given by (\ref{der-connection}) getting 
\bea
& &\lie\bR^d_{\ cab}=\bnb_{[a}\bphi_{b]}P^d_{\ c}+P^d_{\ [b}\bnb_{a]}\bphi_c-P_{c[b}\bnb_{a]}\bphi^d+
\bnb_{[a}\bchi_{b]}\Pi^d_{\ c}+\Pi^d_{\ [b}\bnb_{a]}\bchi_c-\nonumber\\
&-&\Pi_{c[b}\bnb_{a]}\bchi^d+\bphi_{[b}\bnb_{a]}P^d_{\ c}+\bphi_c\bnb_{[a}P^d_{\ b]}-\bphi^d\bnb_{[a}P_{b]c}+
\bchi_{[b}\bnb_{a]}\Pi^d_{\ c}+\bchi_c\bnb_{[a}\Pi^d_{\ b]}-\nonumber\\
&-&\bchi^d\bnb_{[a}\Pi_{b]c}\label{lie-curvatura}.
\eea
The game is now to isolate from this expression $\bnb_a\bphi_b$ and $\bnb_{a}\bchi_b$ (these rank-2 tensors are 
not symmetric in general). 
The forthcoming calculations split in two groups which are dual under the interchange
$P_{ab}\Leftrightarrow\Pi_{ab}$, $p\Leftrightarrow n-p$ (only the calculations for $\bnb_a\bphi_b$ are performed).
Multiplying (\ref{lie-curvatura}) by $P^a_{\ r}$ we obtain
\bea
& &\lie(P^d_{\ r}\bR^r_{\ cab})=P^d_{\ c}\bnb_{[a}\bphi_{b]}+P^d_{\ [b}\bnb_{a]}\bphi_c-
P^d_{\ r}P_{c[b}\bnb_{a]}\bphi^r-P^d_{\ r}\Pi_{c[b}\bnb_{a]}\bchi^r+\nonumber\\
&+&P^d_{\ r}\bphi_{[b}\bnb_{a]}P^r_{\ c}+\bphi_cP^d_{\ r}\bnb_{[a}P^r_{\ b]}-
\bphi^d\bnb_{[a}P_{b]c}+P^d_{\ r}\bchi_{[b}\bnb_{a]}\Pi^r_{\ c}+\bchi_cP^d_{\ r}\bnb_{[a}\Pi^r_{\ b]}.\nonumber\\
\label{seis} 
\eea
Contraction of the indexes $d$-$c$ in the above expression yields
\be
\bnb_a\bphi_b=\bnb_b\bphi_a+\fr{2}{p}\lie(P^d_{\ r}\bR^r_{\ dab}),
\label{nueve}
\ee 
while the contraction of indexes $d-a$ and use of identities (\ref{contraction})-(\ref{PPi}) entails
\be
2\lie(P^d_{\ r}\bR^r_{\ cdb})=P^d_{\ c}\bnb_d\bphi_b+P^d_b\bnb_d\bphi_c-\bphi^r\bnb_rP_{bc}-\bnb_a\bphi^aP_{bc}-
p\bnb_b\bphi_c.
\label{ocho}
\ee
Equations (\ref{nueve}) and (\ref{ocho}) can now be combined in a single expression which is 
\bea
2\lie\left[P^d_{\ r}\bR^r_{\ cdb}-\fr{1}{p}(P^d_{\ c}P^r_{\ q}\bR^q_{\ rdb}+P^d_{\ b}P^r_{\ q}\bR^q_{\ rdc}-
P^r_{\ q}\bR^q_{\ rbc})\right]=\nonumber\\
=(2-p)\bnb_b\bphi_c-\bnb_a\bphi^aP_{bc}-\bphi_d(\bnb_bP^d_{\ c}+\bnb_cP^d_{\ b}+P^{rd}\bnb_rP_{bc}).
\label{diez}
\eea
Multiplying here with $P^{cb}$ leads, after a little bit of algebra, to  
\be
\bnb_a\bphi^a=\fr{1}{1-p}(\lie\bR^0+\phi\bR^0),\ \ p\neq 1,\ \ \bR^0=P^d_{\ r}\bR^r_{\ cdb}P^{cb},
\label{once}
\ee

On the other hand (\ref{diez}) can be further simplified if we take into account the identity
\bnr
& &\bnb_bP^d_{\ c}+\bnb_cP^d_{\ b}-P^{rd}\bnb_rP_{bc}
=\Pi^{rd}\nb_rP_{bc}+\fr{1}{2}(\Pi^{dq}\Pi^r_{\ c}M_{qrb}+\Pi^{dq}\Pi^r_{\ b}M_{qrc})+\\
&+&\fr{1}{2(n-p)}(W_c\Pi^d_{\ b}+W_b\Pi^d_{\ c}),
\enr
which is easily derived by writing all the covariant derivatives with respect to the bi-conformal connection 
of the projectors in terms of ordinary covariant derivatives (proposition \ref{identity-set}). 
So plugging (\ref{once}) into (\ref{diez}) and using previous identity we get
\be
(2-p)\bnb_b\bphi_c=\lie L^0_{bc}+2\bphi^r\bnb_rP_{bc},
\label{normal-1}
\ee
where
\be
L^0_{\ bc}\equiv 
2\left[P^d_{\ r}\bR^r_{\ cdb}-\fr{1}{p}(P^d_{\ c}P^r_{\ q}\bR^q_{\ rdb}+P^d_{\ b}P^r_{\ q}\bR^q_{\ rdc}-
P^r_{\ q}\bR^q_{\ rbc})\right]+\fr{\bR^0}{1-p}P_{bc}.
\label{L0}
\ee  
The duals of (\ref{normal-1}) and (\ref{L0}) are 
\be
(2-n+p)\bnb_b\bchi_c=\lie L^1_{bc}+2\bchi^r\bnb_r\Pi_{bc},
\label{normal-2}
\ee
and 
\bea
L^1_{\ bc}&\equiv & 
2\left[\Pi^d_{\ r}\bR^r_{\ cdb}-\fr{1}{n-p}(\Pi^d_{\ c}\Pi^r_{\ q}\bR^q_{\ rdb}+\Pi^d_{\ b}\Pi^r_{\ q}\bR^q_{\ rdc}-
\Pi^r_{\ q}\bR^q_{\ rbc})\right]+\nonumber\\
+\fr{\bR^1}{1-n+p}\Pi_{bc},&\ \ &\bR^1\equiv\Pi^d_{\ r}\bR^r_{\ cdb}\Pi^{cb}.
\label{L1}
\eea
To complete the normal form we need now the derivatives of $\sphi_a$ and $\schi_a$ which are obtained through 
the differentiation of (\ref{18}) (identity (\ref{lie-conmmutation}) must be used to get these derivatives)
\bea
-p\bnb_b\sphi_a=\lie(\bnb_bE_a)+\fr{1}{2}(\bchi_bE_a+\bchi_aE_b-(\bchi^rE_r)\Pi_{ab})\\
(p-n)\bnb_b\schi_a=\lie(\bnb_bW_a)+\fr{1}{2}(\bphi_bW_a+\bphi_aW_b-(\bphi^rW_r)P_{ab})
\eea

\subsection{Normal form of the differential conditions} 
The above calculations give us the sought normal form for the differential conditions 
(\ref{Bi-conformal}) being these gathered in the  following set of equations  
\bea
(a)\ \ \bnb_a\phi&=&\bphi_a+\sphi_a,\ \bnb_a\chi=\bchi_a+\schi_a,\nonumber\\
(b)\ \ \ \bnb_b\sphi_a&=&\fr{-1}{p}\left[\lie(\bnb_bE_a)+\fr{1}{2}(\bchi_bE_a+\bchi_aE_b-(\bchi^rE_r)\Pi_{ab})\right],\nonumber\\
(c)\ \ \ \bnb_b\schi_a&=&\fr{1}{p-n}\left[\lie(\bnb_bW_a)+\fr{1}{2}(\bphi_bW_a+\bphi_aW_b-(\bphi^rW_r)P_{ab})\right],\nonumber\\
(d)\ \ \ \bnb_b\bphi_c&=&\fr{1}{2-p}\left[\lie L^0_{bc}+2\bphi^r\bnb_rP_{bc}\right],\label{normal-form}\\
(e)\ \ \ \bnb_b\bchi_c&=&\fr{1}{2-n+p}\left[\lie L^1_{bc}+2\bchi^r\bnb_r\Pi_{bc}\right],\nonumber\\
(f)\ \ \ \bnb_b\xi^a&=&\Psi_b^{\ a},\nonumber\\ 
(g)\ \ \ \bnb_b\Psi_c^{\ a}&=&\fr{1}{2}(\bphi_bP^a_{\ c}+\bphi_cP^a_{\ b}-\bphi^aP_{cb}+
\bchi_b\Pi^a_{\ c}+\bchi_c\Pi^a_{\ b}-\bchi^a\Pi_{cb})-\xi^d\bar{R}^a_{\ cdb}.\nonumber 
\eea

A first glance at these equations reveals us that this normal form does not always exist. 
To be precise if either $p=2$ or $p=n-2$ the derivatives $\bnb_a\bphi_b$ and $\bnb_a\bchi_b$ cannot be isolated
and the system cannot be ``closed'' (in fact these derivatives cannot be isolated even if we perform further 
derivatives of any of the above equations). We must also remember at this point that the tensors $L^0_{ab}$ and 
$L^1_{ab}$ are well-defined unless $p=1$ and $p=n-1$ respectively (see equations (\ref{L0}) and (\ref{L1})).
Therefore we have proven the following theorem (see proposition 6.2 of \cite{BI-CONFORMAL})
\begin{theorem}
The only cases in which the Lie algebra $\Gl(S)$ can be infinite dimensional occur if 
and only if $p=1$, $p=2$, $p=n-1$, $p=n-2$.
\end{theorem}
The result of this theorem is intuitively clear if we realize that bi-conformal vector fields are somehow
conformal motions for the projectors $P_{ab}$ and $\Pi_{ab}$. Therefore if any of them projects onto one or 
two dimensional vector spaces the associated Lie algebras may turn out to be infinite dimensional as we have just found.   

Equations (\ref{normal-form}) can fail to exist in the form written above if some of the derivatives involved 
vanish. This happens for instance if part of the gauge functions are constants or their second covariant
 derivatives with respect to the bi-conformal connection   
are zero. In all this work we will assume the conditions for all the derivatives involved in (\ref{normal-form}) 
to exist leaving the study of any other cases for a forthcoming publication. Under these hypotheses bi-conformal
vector fields turn out to be smooth vector fields.

\begin{proposition}
Let $\xiv$ be a bi-conformal vector field at least $C^2$ in a neighbourhood $\U_x$ of a point $x$ belonging to a 
manifold $V$ with a $C^{\infty}$ metric tensor. If $\bnb_b\bphi_a$, $\bnb_b\bchi_a$ are not identically zero on $\U_x$
 then $\xiv\in C^{\infty}(\U_x)$. 
\label{differentiability}
\end{proposition}  
\begin{proof} To prove this result it is enough to show that the covariant derivatives of $\xiv$ with res\-pect to 
the bi-conformal connection exist at any order. The first and second derivatives of $\xiv$ are equations 
(\ref{normal-form})-{\em f} and (\ref{normal-form})-{\em g} and higher derivatives are calculated from 
this last equation. When we derive (\ref{normal-form})-{\em g} we need 
$\bnb_b\bphi_c$ and $\bnb_b\bchi_c$ which by assumption are not zero on $\U_x$ and these are obtained through
equations (\ref{normal-form})-{\em d} and (\ref{normal-form})-{\em e} in which only
derivatives of $\xiv$, $\bphi_a$ and $\bchi_a$ of order less or equal than one appear.
This makes clear that no other equation of (\ref{normal-form}) but the ones mentioned so far are involved 
in the calculation of the derivatives of $\xiv$ and so we can obtain them in as high order as we wish.
 \qed
\end{proof}

Related to this is the following result.
\begin{proposition}
Under the hypotheses of previous proposition if a bi-conformal vector field $\xiv$ is such that 
$\xi^a|_x=0$, $\bnb_b\xi^a|_x=0$, $\bnb_c\bnb_b\xi^a|_x=0$ then $\xiv\equiv 0$ in a neighbourhood of $x$.
\end{proposition}
\begin{proof}
Evaluation of the last equation of (\ref{normal-form}) at $x$ entails
$$
(\bphi_bP^a_{\ c}+\bphi_cP^a_{\ b}-\bphi^aP_{cb}+
\bchi_b\Pi^a_{\ c}+\bchi_c\Pi^a_{\ b}-\bchi^a\Pi_{cb})|_x=0,
$$
from which, projecting down with $P^{ab}$ and $\Pi^{ab}$, we deduce that $\bphi_a|_x=\bchi_a|_x=0$.
Now let $\g(t)$ be a smooth curve on $V$ such that $\g(0)=x$ with $\g(t)$ lying in
a coordinate neighbourhood of $x$ for all $t$ in the interval $(-\epsilon,\epsilon)$. If we denote by $\dot{\g}^a(t)$ the tangent 
vector to this curve we may define the derivatives
$$
\fr{\bar{D}\bphi_a}{dt}\equiv\dot{\g}^r\bnb_r\bphi_a,\ \fr{\bar{D}\bchi_a}{dt}\equiv\dot{\g}^r\bnb_r\bchi_a,\ 
\fr{\bar{D}\xi^a}{dt}\equiv\dot{\g}^r\bnb_r\xi^a,\ \fr{\bar{D}\Psi_c^{\ a}}{dt}\equiv\dot{\g}^r\bnb_r\Psi_c^{\ a}, 
$$   
where all quantities are evaluated on $\g(t)$. Contracting (\ref{normal-form})-{\em d}-(\ref{normal-form})-{\em g} with 
$\dot{\g}^b$ we can transform these equations into a first order ODE system in the variables $\bphi_a(\g(t))$, $\bchi_a(\g(t))$,
$\xi^a(\g(t))$ and $\Psi_a^{\ b}(\g(t))$. From the above all these variables vanish at $t=0$ so according to 
the standard theorem of uniqueness for ODE systems the variables are identically zero along the curve $\g(t)$ (and in particular 
$\xiv|_{\g(t)}=0$). As $\g(t)$ was chosen arbitrarily we conclude that $\xiv\equiv 0$ in a whole neighbourhood of $x$.\qed 
\end{proof}
From the calculations performed in this proof and in the proof of proposition \ref{differentiability}  
we deduce as a simple corollary that 
$\bnb^{(m)}\xi^a|_x\equiv 0$ for all $m\in{\mathbb N}$ if it holds for $m=1,2$.  

\subsection{Constraints}
\label{constraints}
From the above calculation of the normal form, the system variables are read off at once. These are the variables 
appearing under derivation in the left hand side of (\ref{normal-form}). However, they are not algebraically independent
because in the calculation process of (\ref{normal-form}) some of the equations involved do not contain derivatives 
of the variables at all (system constraints). The most evident case of these constraints are the differential 
conditions (\ref{Bi-conformal}) themselves. If we review the whole procedure followed to get (\ref{normal-form}) we 
deduce that the other set of constraints between the system variables is (\ref{18}) so (\ref{normal-form}) must be 
complemented with 
\be
(I)\left.\begin{array}{c}
\lie P_{ab}=\phi P_{ab}\\
\lie\Pi_{ab}=\chi\Pi_{ab}
\end{array}\right\},\ \ 
(II)\left.\begin{array}{l}
\lie E_a=-p\sphi_a\\
\lie W_a=-(n-p)\schi_a\end{array}\right\}
\label{ligaduras}
\ee 
In order to clarify that these two sets of equations are truly the constraints associated to (\ref{normal-form}) 
we must show that they arise as a linear combination of some of the higher covariant derivatives  of 
(\ref{Bi-conformal}) employed to get the normal form (either 
with respect to $\nb$ or $\bnb$). Such higher derivatives are (\ref{9}), (\ref{psi}) which are a linear 
combination of the first derivative of (\ref{Bi-conformal}) (they are constructed through the Lie derivatives of the 
connections $\g^a_{\ bc}$ and $\bbg^a_{\ bc}$) and (\ref{lie-curvature}) which is a linear combination of the first 
derivative of the bi-conformal connection and hence the second derivative of (\ref{Bi-conformal}) (see equations 
(\ref{lie-curvature}), (\ref{lie-connection}) of appendix A). In the case of (\ref{9}) only its projections with 
$P^{ab}$ and $\Pi^{ab}$ (equation \ref{18}) really matter to work out the normal form and these 
are (\ref{ligaduras})-II. As for the other derivatives they do not give rise to 
any more equations with no derivatives of the system variables so the above equations are the only constraints 
we must care about.   

A first application of all the above calculations comes in the following result, 
already proven in \cite{BI-CONFORMAL}  
using a normal form system written in terms of different variables.
\begin{theorem}
If the Lie algebra $\Gl(S)$ is finite dimensional then its dimension is bo\-un\-ded from above by 
$N=p(p+1)/2+(n-p)(n-p+1)/2$.
\label{bounded}
\end{theorem}

\begin{proof}
To prove this theorem we must state what the upper bound to the maximum number of {\em integration constants} for the system 
(\ref{normal-form}) is. As is very well known from the theory of normal systems of P.D.E's (see e. g. \cite{EISENHARTII}) 
such number is the number of system variables minus the number of linearly independent constraints. The following 
table summarises these numbers for our system.

\begin{table}[h]
\caption{Calculation of the highest dimension of $\Gl(S)$}
\label{tab:1}       % Give a unique label
\centering
\begin{tabular}{ccccc|cc}
\hline\hline\noalign{\smallskip}
\multicolumn{5}{c}{system variables} & 
\multicolumn{2}{c}{Constraints}\\
\noalign{\smallskip}\hline\hline\noalign{\smallskip}
$\phi,\chi$ & $\sphi_a$, $\bphi_a$ & $\schi_a$, $\bchi_a$ & $\xi^a$ & $\Psi_a^{\ b}$ & 
eq. (\ref{ligaduras})-I & eq. (\ref{ligaduras})-II\\
\noalign{\smallskip}\hline\noalign{\smallskip}
 2 & $n$ & $n$ & $n$ & $n^2$ & $n(n+1)/2+p(n-p)$ & $n$\\
\end{tabular}
% Or use
%\vspace*{5cm}  % with the correct table height
\end{table}

We have written explicitly the system variables and the total number for each of them. The constraints are also 
indicated together with how many linearly independent equations each constraint amounts to. This last part is not 
evident as opposed to the co\-un\-ting for the system variables so the rest of the proof is devoted to show 
that the numbers given in table \ref{tab:1} for the constraint
equations are indeed correct.  

\medskip
\noindent
{\em Equations (\ref{ligaduras})-I}. First of all, we expand the Lie derivatives of these equations
\be 
\xi^c\bnb_cP_{ab}+\Psi_p^{\ c}(\dd^p_{\ a}P_{bc}+\dd^p_{\ b}P_{ac})=\phi P_{ab},\ \ 
\xi^c\bnb_c\Pi_{ab}+\Psi_p^{\ c}(\dd^p_{\ a}\Pi_{bc}+\dd^p_{\ b}\Pi_{ac})=\chi\Pi_{ab},\ \
\label{expansion}
\ee
where the standard definition of the Lie derivative of a tensor $P_{ab}$ has been applied
\be
\lie P_{ab}\equiv\xi^c\bnb_c P_{ab}+\bnb_a\xi^cP_{cb}+\bnb_b\xi^cP_{ac},
\ee   
(observe that the general formula of the Lie derivative with respect to a vector in terms of its covariant
derivatives still holds under an symmetric connection). We define the new indexes 
$A$, $B$, $B'$ in such a way that
$$
P_A\equiv P_{a b},\ \Psi_{B}\equiv\bnb_p\xi^q,\ \xi^{B'}=\xi^c, 
$$
so capital indexes group together certain combinations of small indexes (explicitly $A=\{a,b\}$, $B=\{p, q\}$ and
$B'=c$. The ranges of the new indexes are 
$A=1,\dots, n(n+1)/2$, $B=1,\dots n^2$, $B'=1,\dots n$ (note that the above expressions are symmetric in $a, b$).
Using these new labels we can write in
matrix notation the homogeneous system posed by these constraints (we only concentrate in 
the first of (\ref{expansion}))
\be
(\begin{array}{ccc}
   M_A^{\ B} & (\bnb P)_{AB'} & P_A 
\end{array}) \left(\begin{array}{c}
                              \Psi_B\\
                              \xi^{B'}\\
                              -\phi \end{array}\right)=0, 
\label{matrix-form}
\ee
($A$= row index, $B$, $B'$=column indexes) 
where the explicit expressions of the matrices read
$$
\begin{array}{r}
(\bnb P)_{AB'}=\bnb_cP_{ab},\\
\overline{A\ }\ \end{array}\ \begin{array}{rr}
\ M_A^{\ B}=\dd^p_{\ a}P_{bc} &\!+\dd^p_{\ b}P_{ac}\\
            \overline{\ A\ }\ \ &\overline{\ A\ \ }\ \ \end{array}.
$$

The number of linearly independent equations is just the rank of the 
matrix system of (\ref{matrix-form}). 
In principle the rank of this matrix will depend on the projector $P_{ab}$ and its covariant 
derivative, 
meaning this that it will depend on the geometry of the manifold. However, since we are interested in spaces with a 
maximum 
number of bi-conformal vector fields it is enough to find the least rank of the above matrix for all possible 
projectors $P_{ab}$. We start first studying the rank of $M_A^{\ B}$ whose  
 nonvanishing components occur in the following cases (no summation 
over the repeated indexes)
\begin{eqnarray*}
p=a,\ c=b &\Rightarrow & \left\{\begin{array}{cc}
M_Q^{\ Q}=\dd^a_{\ a}P_{bb}, & a\neq b,\\
M_Q^{\ Q}=2 P_{aa},\ a=b. & 
       \end{array}\right.\\
p=b,\ c=a &\Rightarrow &\
 M_Q^{\ Q'}=\dd^b_{\ b}P_{aa},\ a\neq b,
\end{eqnarray*}
where we have assumed that we are working in the common (orthonormal) basis of eigenvectors of $P^a_{\ b}$ and 
$\Pi^a_{\ b}$ so 
$$
P^{a}_{\ b}=\mbox{diag}(\overbrace{1\dots 1}^{p} 0 \dots 0),\  
\Pi^a_{\ b}=\mbox{diag}(0\dots 0\overbrace{1\dots 1}^{n-p}).
$$ 
Hence we only need to count how many components of the type $M_{Q}^{\ Q}$ are different from zero 
because by construction these elements give rise to linearly independent rows of the matrix $M_A^{\ B}$
(the elements $M_{Q}^{\ Q'}$ are in the same row of the matrix $M_A^{\ B}$ and they do not increase its rank).
The sought number can be obtained from the following diagram
gathering into blocks the $A$ indexes of the rows containing non-zero
elements (we express each index in terms of tensor indexes following the notation $A=(a,b)$) 
$$
\begin{array}{l|l}
\mbox{Block 1}=\{\overbrace{(1,1)\ (1,2)\ \dots\ (1,p)}^{p}\} &\ \ \mbox{Block $p+1$}=\{\overbrace{(p+1,1)\ \dots (p+1,p)}^{p}\}\\
\mbox{Block 2}=\{\overbrace{(2,2)\ (2,3)\ \dots\ (2,p)}^{p-1}\}&\ \ \mbox{Block $p+2$}=\{\overbrace{(p+2,1)\ \dots\ (p+2,p)}^p\}\\  
\mbox{Block 3}=\{\overbrace{(3,3)\ (3,4)\ \dots\ (3,p)}^{p-2}\}&\ \  \mbox{Block $p+3$}=\{\overbrace{(p+3,1)\ \dots\ (p+3,p)}^p\}\\   
\dots\dots\dots\dots\dots\dots\dots\dots\dots\dots\dots&\ \ \dots\dots\dots\dots\dots\dots\dots\dots\dots\dots\dots\dots\dots \\
\mbox{Block $p$}=\{\overbrace{(p,p)}^1\} &\ \  \mbox{Block $p+n$}=\{\overbrace{(n,1)\ \dots\ (n,p)}^p\},
\end{array}
$$
from which the rank of $M_A^{\ B}$ is 
$$
1+\dots+p+p(n-p)=\fr{1}{2}p(p+1)+p(n-p).
$$
Notice that this rank only depend on algebraic properties of the projector $P_{ab}$ and not on its actual form at some 
concrete space. Addition of the matrices $\bnb_AP^{B'}$ and $P_A$ only increase the rank of the matrix of the homogeneous system 
(\ref{matrix-form}) and so we do not need to take them into account.  The total 
number of constraints posed by (\ref{ligaduras})-I is then the rank of $M_A^{\ B}$ plus the rank of the matrix $N_A^{\ B}$ constructed
replacing $P_{ab}$ by $\Pi_{ab}$
\bnr
rank(M)&+&rank(N)=\fr{1}{2}p(p+1)+p(n-p)+\fr{1}{2}(n-p)(n-p+1)+p(n-p)=\\
&=& \fr{1}{2}n(n+1)+p(n-p).
\enr 

\medskip
\noindent
{\em Equations (\ref{ligaduras})-II.} In order to perform the analysis of these constraints it is enough to realize that 
the 1-forms $\sphi_a$ and $\schi_a$ appearing in the right hand side of each equation are invariant under the projectors 
$\Pi^a_{\ b}$ and $P^a_{\ b}$ respectively. Therefore each constraint contains at least $n-p$ and $p$ linearly independent 
equations being $n$ the total sum of them.

The upper bound $N$ is then 
\bnr
N=2+n+n+n+n^2-(n+\fr{1}{2}n(n+1)+p(n-p))&=&\\
=\fr{1}{2}(p+1)(p+2)&+&\fr{1}{2}(n-p+1)(n-p+2).
\enr
\qed
\end{proof}

\begin{remark}
This proof does not guarantee the existence of a Lie algebra $\Gl(S)$ in which the dimension $N$ is attained. However in 
section \ref{sufficient} we will exhibit explicit examples of a pseudo-Riemannian spaces possessing $N$ linearly independent 
bi-conformal vector fields deriving also a geometric characterization of these spaces.    
\end{remark}

\section{First integrability conditions}
\label{first-integrability}
Once the normal form (\ref{normal-form}) has been obtained we must study next its integrability conditions. These are geometric 
conditions arising from the consistency between the covariant derivatives commutation rules and the expressions for these 
derivatives given by (\ref{normal-form}). In this paper we are only interested in the {\em first integrability conditions} so 
we will concentrate next in compatibility conditions obtained from the commutation of two covariant derivatives (Ricci 
identity)
\bnr
\bnb_a\bnb_b\Xi^{a_1\dots a_r}_{\ b_1\dots b_s}-\bnb_b\bnb_a\Xi^{a_1\dots a_r}_{\ b_1\dots b_s}&=&\\
=\sum_{q=1}^r\bR^{a_q}_{\ tab}\Xi^{a_1\dots a_{q-1}ta_{q+1}\dots a_r}_{\ b_1\dots b_s}&-&
\sum_{q=1}^{s}\bR^t_{\ b_qab}\Xi^{a_1\dots a_r}_{\ b_1\dots b_{q-1}tb_{q+1}\dots b_s},
\enr
where we must replace the tensor $\Xi^{a_1\dots a_r}_{\ b_1\dots b_s}$ by the system variables and apply 
(\ref{normal-form}) to work out the covariant derivatives. Each one of the equations derived in this fashion only 
involves system variables and is called {\em integrability condition}. These integrability conditions can be further
covariantly differentiated yielding integrability conditions of higher degree. The constraint equations (\ref{ligaduras}) also 
give rise to integrability conditions when differentiated in the obvious way. 

We are going next to work out the full set of first integrability conditions of the normal form (\ref{normal-form}) and 
the constraints (\ref{ligaduras}). Part of these calculations were already performed in \cite{BI-CONFORMAL} but we 
will repeat everything in order to make the work presented in this section self-contained. The order followed to work out 
the integrability conditions of each equation of (\ref{normal-form}) will not coincide with the order followed in their 
presentation.  

The simplest integrability condition is that of equation (\ref{normal-form})-{\em f} 
$$
\bnb_c\bnb_b\xi^a-\bnb_b\bnb_c\xi^a=\bR^a_{\ rcb}\xi^r=\bnb_c\Psi_b^{\ a}-\bnb_b\Psi_c^{\ a},
$$
being this expression an identity as is easily checked by replacing the covariant derivatives of $\Psi_a^{\ b}$. Next we 
tackle the integrability conditions of (\ref{normal-form})-{\em g} which in fact are given by (\ref{lie-curvatura}) with all the 
covariant derivatives of $\bphi_a$ and $\bchi_a$ replaced by their values
(\ref{normal-form})-{\em d} and (\ref{normal-form})-{\em e}. To see this just recall that (\ref{lie-curvatura}) was obtained 
through identity (\ref{lie-curvature}) which can be rewritten as
$$
\bnb_a\bnb_b\Psi_c^{\ d}-\bnb_b\bnb_a\Psi_c^{\ d}=\lie\bR^d_{\ cab}-
\Psi_a^{\ r}\bR^d_{\ crb}+\Psi_b^{\ r}\bR^d_{\ cra}-\xi^r\bnb_r\bR^d_{\ cab},
$$
once the Lie derivatives of the connection (identity (\ref{lie-xi}) have been substituted. Clearly the direct substitution in 
(\ref{lie-curvatura}) results in a rather long expression so it is better to perform the replacements separately in 
(\ref{seis}) and its counterpart with the interchanges $P_{ab}\leftrightarrow\Pi_{ab}$, etc. Adding the two equations
so obtained 
yields after lengthy algebra
\bea
\fr{1}{2}\lie T^d_{\ cab}=\fr{\bphi_r}{2-p}(P^d_{\ [b}\La^r_{\ a]c}&+&P^d_{\ q}\Up^{rq}_{\ [b}P_{a]c})+\fr{\bchi_r}{2-(n-p)}
(\Pi^d_{\ [b}\bLa^r_{\ a]c}+\Pi^d_{\ q}\bUp^{rq}_{\ [b}P_{a]c})+\nonumber\\
+\bphi_{[b}\bnb_{a]}P^d_{\ c}+\bphi_c\bnb_{[a}P^d_{\ b]}&+&\bphi^d\bnb_{[b}P_{a]c}+\bchi^d\bnb_{[b}\Pi_{a]c}+
\bchi_{[b}\bnb_{a]}\Pi^d_{\ c}+\bchi_c\bnb_{[a}\Pi^d_{\ b]},\nonumber\\
\ 
\label{cond-g}
\eea   
where by convenience we introduce the tensors
\bnr
\La^d_{\ bc}&\equiv&2P^{dr}\bnb_rP_{bc},\ \bLa^d_{\ bc}=2\Pi^{dr}\bnb_r\Pi_{bc},\\
\Up^{sc}_{\ b}&\equiv&2P^{sr}P^{cq}\bnb_rP_{qb}+(2-p)\bnb_bP^{sc},\ 
\bUp^{sc}_{\ b}\equiv2\Pi^{sr}\Pi^{cq}\bnb_r\Pi_{qb}+(2-n+p)\bnb_b\Pi^{sc},
\enr 
and the tensor $T^d_{\ cab}$ is given by
\bea
T^d_{\ cab}\equiv 2\bR^d_{\ cab}-\fr{2}{2-p}(P^d_{\ c}L^0_{[ab]}+P^d_{\ [b}L^0_{a]c}+P_{c[a}L^0_{b]q}P^{qd})+
\nonumber\\
-\fr{2}{2-n+p}(\Pi^d_{\ c}L^1_{[ab]}+\Pi^d_{\ [b}L^1_{a]c}+\Pi_{c[a}L^1_{b]q}\Pi^{qd}).
\label{tensor-tt}
\eea
This tensor will play an important role as will be seen later.

An interesting invariance property of some of the above tensors needed in future calculations is
\be
\lie\La^d_{\ bc}=0,\ \ \lie\bLa^d_{\ bc}=0
\label{inv-3}
\ee
which are easily obtained from (\ref{cond-ligaduras-up}).

Next we address the first integrability conditions of (\ref{normal-form})-{\em b} and (\ref{normal-form})-{\em c}. Again the 
calculations are tedious but straightforward (formula (\ref{lie-conmmutation}) is used to commute the Lie derivative and the 
covariant derivative when differentiating both equations and the covariant derivatives of $\bphi_a$ and $\bchi_a$ 
are calculated through (\ref{normal-form})-{\em d} and (\ref{normal-form})-{\em e})
\bea
& &E_d\lie(\Pi^d_{\ r}T^r_{\ acb})=\bchi_a\bnb_{[c}E_{b]}+\bchi_{[b}\bnb_{c]}E_a-\bchi^r\bnb_{[c}(\Pi_{b]a}E_r)+\nonumber\\
&+&(P^r_{\ a}\bphi_{[c}+\bphi_aP^r_{\ [c}-\bphi^rP_{a[c}+
\Pi^r_{\ a}\bchi_{[c}+\bchi_a\Pi^r_{\ [c}-\bchi^r\Pi_{a[c})\bnb_{b]}E_r+\nonumber\\
&+&\fr{1}{2-n+p}\bchi_qE_r\bUp^{qr}_{\ [b}\Pi_{c]a},\label{cond-b}\\
& &W_d\lie(P^d_{\ r}T^r_{\ acb})=\bphi_a\bnb_{[c}W_{b]}+\bphi_{[b}\bnb_{c]}W_a-\bphi^r\bnb_{[c}(P_{b]a}W_r)+\nonumber\\
&+&(\Pi^r_{\ a}\bchi_{[c}+\bchi_a\Pi^r_{\ [c}-\bchi^r\Pi_{a[c}+
P^r_{\ a}\bphi_{[c}+\bphi_aP^r_{\ [c}-\bphi^rP_{a[c})\bnb_{b]}W_r+\nonumber\\
&+&\fr{1}{2-p}\bphi_qW_r\Up^{qr}_{\ [b}P_{c]a}.\label{cond-c}
\eea
The first integrability conditions of (\ref{normal-form})-{\em a} are easier to handle
\bnr
0=\bnb_a\bnb_b\phi-\bnb_b\bnb_a\phi=\bnb_a\bphi_b-\bnb_b\bphi_a+\bnb_a\sphi_b-\bnb_b\sphi_a\\
0=\bnb_a\bnb_b\chi-\bnb_b\bnb_a\chi=\bnb_a\bchi_b-\bnb_b\bchi_a+\bnb_a\schi_b-\bnb_b\schi_a,\\
\enr
from which we readily obtain by means of (\ref{normal-form})-{\em b}, (\ref{normal-form})-{\em c}, (\ref{normal-form})-{\em d},
(\ref{normal-form})-{\em e}
\bea
\lie\left(\fr{1}{2-p}L^0_{[ab]}-\fr{1}{p}\bnb_{[a}E_{b]}\right)=0,\label{cond-a1}\\
\lie\left(\fr{1}{2-n+p}L^1_{[ab]}-\fr{1}{n-p}\bnb_{[a}W_{b]}\right)=0.\label{cond-a2}
\eea
These are in principle independent integrability conditions. However, several calculations in explicit examples 
always have shown that the tensors inside the rounded brackets vanish identically so we believe that (\ref{cond-a1}) and 
(\ref{cond-a2}) are in fact identities holding for any vector field $\xiv$. The explicit proof of this has not been completed
yet.

Finally only the first integrability conditions of (\ref{normal-form})-{\em d} and (\ref{normal-form})-{\em e} are left.
The calculation method is similar to the above cases and the result is (equation (\ref{inv-3}) is used along the way)
\bea
\fr{2-p}{2}\bphi_rT^r_{\ ceb}&=&2\lie\left(\bnb_{[e}L^0_{b]c}+\fr{1}{2-p}\La^d_{\ c[b}L^0_{e]d}\right)+
\bchi_{[e}L^0_{b]q}\Pi^q_{\ c}+\nonumber\\
+\bchi_c\Pi^q_{\ [e}L^0_{b]q}-\bchi^q\Pi_{c[e}L^0_{b]q}
&+&\fr{2}{2-p}\bphi_r(\La^d_{c[b}\La^r_{e]d}+(2-p)\bnb_{[e}\La^r_{\ b]c}),\label{cond-d}\\
\fr{2-n+p}{2}\bchi_rT^r_{\ ceb}&=&2\lie\left(\bnb_{[e}L^1_{b]c}+\fr{1}{2-n-p}\bLa^d_{\ c[b}L^1_{e]d}\right)+
\bchi_{[e}L^1_{b]q}P^q_{\ c}+\nonumber\\
+\bphi_cP^q_{\ [e}L^1_{b]q}-\bphi^qP_{c[e}L^1_{b]q}&+&
\fr{2}{2-n+p}\bchi_r(\bLa^d_{c[b}\bLa^r_{e]d}+(2-n+p)\bnb_{[e}\bLa^r_{\ b]c}).\label{cond-e} 
\eea
 
The integrability conditions of the constraints (\ref{ligaduras}) result from their covariant derivative. We only need to take 
care of (\ref{ligaduras})-I because the differentiation of (\ref{ligaduras})-II results in (\ref{normal-form})-{\em b} and 
(\ref{normal-form})-{\em c} which are part of the normal form. Differentiation of these equations with respect to $\bnb$
yield after some algebra 
\be
\lie\bnb_cP_{ab}=\phi\bnb_cP_{ab}+\sphi_cP_{ab},\ \ 
\lie\bnb_c\Pi_{ab}=\chi\bnb_c\Pi_{ab}+\schi_c\Pi_{ab}.
\label{cond-ligaduras-down}
\ee
For completeness we provide also these equations with the $ab$ indexes raised
\be
\lie\bnb_cP^{ab}=-\sphi_cP^{ab}-\phi\bnb_cP^{ab},\ \ 
\lie\bnb_c\Pi^{ab}=-\schi_c\Pi^{ab}-\chi\bnb_c\Pi^{ab},\ \ 
\label{cond-ligaduras-up}
\ee
and the invariance laws
\be
\lie\bnb_cP^a_{\ b}=0,\ \ \lie\bnb_c\Pi^a_{\ b}=0.
\label{mixed-invariance}
\ee
These equations close the whole suite of first integrability conditions. We will obtain geometric information from these
conditions in the next section but before doing that let us show that likewise other symmetry transformations studied 
in Differential Geometry we can find {\em geometric invariants} associated to bi-conformal vector fields.

\subsection{Geometric invariants} 
Equations (\ref{cond-g})-(\ref{mixed-invariance}) look rather more complicated than the first integrability conditions of 
other symmetries such as isometries or conformal motions. There is though a common point between all of them and it is 
the existence of {\em geometric invariants}, namely, tensors whose Lie derivative with respect to any vector field 
generating the symmetry under study vanishes. Here we face a richer set of invariants than in the isometric or conformal 
case as we are able to construct continuous multi-parameter families of invariants.

\begin{theorem}
The six-parameter family of rank-3 tensors given by 
\bnr
W^a_{\ bc}(\l_1,\l_2,\l_3;\mu_1,\mu_2,\mu_3)&\equiv&\l_1\La^a_{\ bc}+\l_2\bnb_cP^a_{\ b}+\l_3\bnb_bP^a_{\ c}+
\mu_1\bLa^a_{\ bc}+\\
&+&\mu_2\bnb_c\Pi^a_{\ b}+\mu_3\bnb_b\Pi^a_{\ c},
\enr
is a geometric invariant.
\label{six-parameter}
\end{theorem}
\begin{proof}
This is straightforward from (\ref{inv-3}) and (\ref{mixed-invariance}).\qed
\end{proof}
These invariants are not the only ones which can be found for bi-conformal vector fields.  
\begin{theorem}
Let $\ ^{||}C^d_{\ cab}$ and $\ ^{\perp}C^d_{\ cab}$ be the tensors constructed from $T^d_{\ cab}$ by 
the fo\-llo\-wing definitions
\be
\ ^{||}C^d_{\ cab}\equiv P^d_{\ r}P^s_{\ c}P^t_{\ a}P^q_{\ b}T^r_{\ stq},\  
\ ^{\perp}C^d_{\ cab}\equiv\Pi^d_{\ r}\Pi^s_{\ c}\Pi^t_{\ a}\Pi^q_{\ b}T^r_{\ stq}.
\label{conformal}
\ee
Then, for any bi-conformal vector field $\xiv$ we have the family of invariance rules 
\be
\lie(\l_1\ ^{||}C^d_{\ cab}+\l_2\ ^{\perp}C^d_{\ cab})=0,\ \forall\l_1,\l_2\in\r.
\ee
\label{geometric-invariants}
\end{theorem}
\begin{proof}
It is enough to show that the Lie derivative of $\ ^{\perp}C^d_{\ cab}$ and $\ ^{||}C^d_{ cab}$ vanishes. 
The proof of this result is accomplished by projecting equations (\ref{cond-g}) in all its indexes 
with the projectors $P_{ab}$ and $\Pi_{ab}$ respectively and realising that the right hand side is zero. In fact 
all the terms of the right hand side of (\ref{cond-g}) are zero when full projected in all their indexes with either 
$P_{ab}$ or $\Pi_{ab}$. To see this we note that the simple result
$$
P^r_{\ a}P^s_{\ b}P^c_{\ q}\nb_rP^q_{\ s}=0,\ \Pi^r_{\ a}\Pi^s_{\ b}\Pi^c_{\ q}\nb_r\Pi^q_{\ s}=0, 
$$
entails
$$
P^r_{\ a}P^s_{\ b}P^c_{\ q}\bnb_rP^q_{\ s}=0,\ \Pi^r_{\ a}\Pi^s_{\ b}\Pi^c_{\ q}\bnb_r\Pi^q_{\ s}=0, 
$$
as can be explicitly checked from identities (\ref{identity-1})-(\ref{identity-3}) (this result is still true 
even if we raise or lower any index of the projectors). From this last equation it is now 
easy to prove the asserted claim for all of the right hand side terms of (\ref{cond-g}).\qed 
\end{proof}
Therefore the diffeomorphisms generated by bi-conformal vector fields keep invariant
all the afore-mentioned families of tensors. Some of 
these results have immediate geometric consequences. If $\{\f_s\}_{s\in\r}$ is a one-parameter group of bi-conformal transformations
then the invariance laws of the tensors  $\ ^{||}C^a_{\ bcd}$ and $\ ^{\perp}C^a_{\ bcd}$ are equivalent to
 (we use index-free notation)
\be
({}^{*}\f_s^*{}^{||}\C)(\G,\bP)={}^{||}\C(\G,\bP),\ \ ({}^{*}\f_s^*{}^{\perp}\C)(\G,\bPi)={}^{\perp}\C(\G,\bPi),
\label{finite}
\ee
where by ${}^{||}\C(\G,\bP)$ and ${}^{\perp}\C(\G,\bPi)$ we mean the 
tensors $\pC$ and $\oC$ calculated from the metric $\G$ and the orthogonal complementary projectors $\bP$, $\bPi$.
This equation is still true for any bi-conformal transformation $\Phi$ even if it does not belong to any one-parameter group 
of bi-conformal transformations (see appendix B).
On the other hand the formula
\be
({}^*\Phi^*{}^{||}\C)(\G,\bP)={}^{||}\C(\Phi^*\G,\Phi^*\bP),\ \ 
({}^*\Phi^*{}^{\perp}\C)(\G,\bPi)={}^{\perp}\C(\Phi^*\G,\Phi^*\bPi), 
\label{Phi}
\ee
holds for any diffeomorphism $\Phi:V\rightarrow V$. 
Therefore combination of (\ref{finite}) and (\ref{Phi}) yields the geometric property
\be
\pC(\alpha_1\bP+\alpha_2\bPi,\alpha_1\bP)=\pC(\G,\boldsymbol{P}),\ \ 
\oC(\alpha_1\bP+\alpha_2\bPi,\alpha_2\bPi)=\oC(\G,\boldsymbol{\Pi}),\ \ 
\label{property}
\ee
which is true for any bi-conformal relation $\Phi:V\rightarrow V$ with $\Phi^*\bP=\alpha_1\bP$, $\Phi^*\bPi=\alpha_2\bPi$.
Moreover this is true for any pair of positive functions $\alpha_1$, $\alpha_2$ even if they do not arise from 
any bi-conformal transformation (see appendix B for a proof of this).
Properties (\ref{finite})-(\ref{property}) still hold if we replace $\pC$, $\oC$ by any member of the family 
of tensors introduced in theorems \ref{six-parameter} and \ref{geometric-invariants}. 
\begin{theorem}
If $\boldsymbol{P}$ and $\bPi$ are orthogonal complementary projectors with respect to a flat pseudo-Riemannian metric 
$\bfeta$ then for any $\alpha_1$, $\alpha_2$ $\in C^2$ 
$$
\pC(\alpha_1\boldsymbol{P}+\alpha_2\boldsymbol{\Pi},\alpha_1\bP)=0,\ \ 
\oC(\alpha_1\boldsymbol{P}+\alpha_2\boldsymbol{\Pi},\alpha_2\bPi)=0.  
$$
\label{ppC}
\end{theorem}
\begin{proof}
This follows straight away from (\ref{property}) because trivially $\pC(\bfeta,\bP)=\oC(\bfeta,\bPi)=0$
for any pair of orthogonal complementary projectors $\bP$, $\bPi$.
\qed
\end{proof}

We will prove in section \ref{sufficient}
that the converse of this theorem is also true, that is to say if both $\pC(\G,\bP)$ and 
$\oC(\G,\bPi)$ vanish for a certain metric $\G=\bP+\bPi$ then $\bP$ and $\bPi$ are conformal to orthogonal and complementary 
projectors with respect to a certain flat metric. Moreover we suspect (see example \ref{example2}) that the vanishing 
of $\pC(\G,\bP)$ ($\oC(\G,\bPi)$) alone is a necessary and sufficient condition for the existence of conformally flat 
foliations on $(V,\rmg_{ab})$ being $P_{ab}$ $(\Pi_{ab})$ the metric tensor of the hypersurfaces forming these foliations.
The rigorous proof of this statement is under investigation. 

\section{Complete integrability}
\label{complete}
Our next task is to decide when first integrability conditions presented in previous sec\-ti\-on become a set of identities
for every election of the independent variables of the normal form (\ref{normal-form}). 
Under such circumstances the Lie algebra of bi-conformal vector fields attains its greatest dimension $N$ and so if we are
able to find spaces $(V,\G)$ in which this happens we will have proven that the bound $N$ is actually reached 
(see \cite{EISENHARTII}).
The first problem which 
we come across to is that not all the variables appearing in (\ref{normal-form}) are independent as there are constraints 
which must be taken into account. However, some of the variables of (\ref{normal-form}) are not involved in the constraint
equations (\ref{ligaduras}) and this fact will allow us to find necessary and sufficient geometric conditions for the 
whole set of first integrability conditions to become identities. The variables which are not constrained by 
(\ref{ligaduras}) are $\bphi_a$ and $\bchi_a$ so we will separate out in the integrability conditions all the contributions 
involving these variables and derive the geometric implications. We will perform next this procedure step by step for each of
the integrability conditions obtained before (actually we will not need to analyse all the conditions as some of them 
will turn into identities if the geometric conditions entailed by others are imposed). The full calculation is rather cumbersome
and only its main excerpts will be shown so the reader interested in the geometric conditions characterizing 
the complete integrability should jump straight to theorem 
\ref{complete-integrability}. 

\medskip
\noindent
{\em Equation (\ref{cond-g})}. This equation can be rewritten as 
\be
\lie T^d_{\ cab}=\bphi_rM^{rd}_{\ cab}+\bchi_rN^{rd}_{\ cab}, 
\label{cond-g1}
\ee  
where
\bea
M^{rd}_{\ cab}&\equiv&\fr{2}{2-p}P^r_{\ s}(\La^s_{\ c[a}P^d_{\ b]}+\Up^{sd}_{\ [b}P_{a]c})+2P^r_{\ [b}\bnb_{a]}P^d_{\ c}+
2P^r_{\ c}\bnb_{[a}P^d_{\ b]}+2P^{dr}\bnb_{[b}P_{a]c},\nonumber\\
\label{def-M}\\
N^{rd}_{\ cab}&\equiv&\fr{2}{2-n+p}\Pi^r_{\ s}(\bLa^s_{\ c[a}\Pi^d_{\ b]}+\bUp^{sd}_{\ [b}\Pi_{a]c})+2\Pi^r_{\ [b}\bnb_{a]}\Pi^d_{\ c}+
2\Pi^r_{\ c}\bnb_{[a}\Pi^d_{\ b]}+\nonumber\\
&+&2\Pi^{dr}\bnb_{[b}\Pi_{a]c}.\label{def-N}
\eea
As (\ref{cond-g1}) must be true for every $\bphi_a$, $\bchi_a$ we have
\be
P^r_{\ s}M^{sd}_{\ cab}=M^{rd}_{\ cab}=0,\ \ \Pi^r_{\ s}N^{sd}_{\ cab}=N^{rd}_{\ cab}=0.
\label{cuarenta}
\ee
Let us study these geometric conditions (it is enough to concentrate on the first condition because the 
other is dual and comes through the usual replacements). Contracting the indexes $d$ with $b$ in this equation we get
\be
\fr{2}{2-p}[pP^{rq}\bnb_qP_{ac}-P^{rq}(P_a^{\ s}\bnb_qP_{sc}+P_c^{\ s}\bnb_qP_{sa})]+P^{sr}\bnb_sP_{ac}=0\label{d-b}.
\ee
Projecting this in the $a$-index with $P_a^{\ s}$ gives the condition
\be
P_a^{\ s}P^{rq}\bnb_qP_{sc}=0,
\ee
which put back in (\ref{d-b}) yields
\be
P^{rs}\bnb_{s}P_{ac}=0\Rightarrow\La^r_{\ ac}=0,\ \ \Up^{sc}_{\ b}=(2-p)\bnb_bP^{sc}.
\label{1st-condition}
\ee
Now using this result we contract $r$ with $b$ in the first of (\ref{cuarenta})
\be
(p-1)\bnb_aP^d_{\ c}-P_a^{\ s}\bnb_sP^d_{\ c}-P_c^{\ s}\bnb_sP^d_{\ a}+P_c^{\ s}\bnb_aP^d_{\ s}=0,
\label{r-b}
\ee
from which we get
\be
P_a^{\ q}P_c^{\ s}\bnb_qP^d_{\ s}=0.
\ee
Combining this with (\ref{r-b}) we readily obtain
\be
\bnb_aP^d_{\ c}=0.
\label{cqe}
\ee
Finally multiplying by $P^{ac}$ in the first of (\ref{cuarenta}) gives
\be
P^r_{\ s}(-P^a_{\ b}\bnb_aP^{sd}+p\bnb_bP^{sd})-P^{dr}E_b=0.
\label{zzz}
\ee
Projection of this equation in the index $b$ leads to 
$$
P^a_{\ b}P^r_{\ s}\bnb_aP^{sd}=0,
$$
which combined with (\ref{zzz}) implies 
\be
\bnb_bP^{rd}=\fr{1}{p}E_bP^{rd}.
\label{pre}
\ee
This last condition can be written with the indexes of the projector lowered if we use (\ref{cqe})
\be
\bnb_bP_{cd}=-\fr{1}{p}E_bP_{dc}.
\label{lll}
\ee
The dual conditions for the projectors $\Pi^{ab}$, $\Pi_{ab}$ coming from the vanishing of (\ref{def-N}) are
\be
\bnb_c\Pi^{ab}=\fr{1}{n-p}W_c\Pi^{ab},\ \bnb_c\Pi^a_{\ b}=0, \bnb_c\Pi_{ab}=-\fr{1}{p}W_c\Pi_{ab}.
\label{qrs}
\ee
Conversely if equations (\ref{cqe}), (\ref{pre}), (\ref{lll}) and (\ref{qrs}) are assumed a simple calculation 
tells us that the tensors defined by (\ref{def-M}) and (\ref{def-N}) vanish. In fact all the above geometric 
conditions can be combined in a single simpler expression.
\begin{proposition}
The following assertion is true
\be
T_{abc}=0\Longleftrightarrow
\bnb_aP_{bc}=-\fr{1}{p}E_aP_{bc},\ \bnb_a\Pi_{bc}=-\fr{1}{n-p}W_a\Pi_{bc}.
\label{assertion}
\ee
\label{ftp}
\end{proposition} 

\begin{proof}
First of all, it is convenient to rewrite the condition $T_{abc}=0$ in an appropriate form.
 From (\ref{tensor-t}) we have 
\be
M_{abc}=\fr{1}{p}E_aP_{bc}-\fr{1}{n-p}W_a\Pi_{bc}.
\label{star}
\ee
Using the definition $M_{abc}=\nb_bP_{ac}+\nb_cP_{ab}-\nb_aP_{bc}$ we can isolate $\nb_bP_{ac}$ getting
\be
\nb_bP_{ac}=\fr{1}{2p}(E_aP_{bc}+E_cP_{ab})-\fr{1}{2(n-p)}(W_a\Pi_{bc}+W_c\Pi_{ba}).
\label{25}
\ee
Each (\ref{star}) and (\ref{25}) are equivalent to $T_{abc}=0$. Next we show the equivalence of $T_{abc}=0$ to 
the conditions (\ref{assertion}). Expanding 
$\bnb_aP_{bc}$ by means of (\ref{identity-1}) and use of the conditions upon the covariant derivatives 
yields
\bea
\nb_bP_{ac}&=&\fr{1}{2p}(E_aP_{bc}+E_cP_{ab})+\fr{1}{2}(P_{cp}M^p_{\ ba}+P_{ap}M^p_{\ bc})\\
\nb_b\Pi_{ac}&=&\fr{1}{2(n-p)}(W_a\Pi_{bc}+W_c\Pi_{ab})-\fr{1}{2}(\Pi_{cp}M^p_{\ ba}+\Pi_{ap}M^p_{\ bc}),
\eea
which are equivalent to (write $M_{abc}$ in terms of $\nb_aP_{bc}$ and $\nb_a\Pi_{bc}$ respectively)
\be
P_{cp}M^p_{\ ab}=-\fr{1}{n-p}W_c\Pi_{ab},\ \ \Pi_{cp}M^p_{\ ab}=\fr{1}{p}E_cP_{ab},
\ee
whose addition leads to $T_{abc}=0$. Conversely, suppose that $T_{abc}=0$. Then inserting (\ref{star}) and 
(\ref{25}) into (\ref{identity-1}) gives us the condition on $\bnb_aP_{bc}$ at once.
 The calculation for $\bnb_a\Pi_{bc}$ is similar using the identities written in terms of $\Pi_{ab}$.\qed
\end{proof}
\begin{proposition}
$$
T_{abc}=0\Longrightarrow \bnb_cP^a_{\ b}=0.
$$
\label{up-down}
\end{proposition}

\begin{proof}
To prove this we use identity 
(\ref{identity-2}) and replace $M_{abc}$ by the expression found in (\ref{star})
\be
\bnb_aP^b_{\ c}=\nb_aP^b_{\ c}-\fr{1}{2p}(E^bP_{ac}+E_cP^b_{\ a})+\fr{1}{2(n-p)}(W_c\Pi^b_{\ a}+W^b\Pi_{ac}),
\ee
which vanishes by (\ref{25}).\qed
\end{proof}

\begin{remark}
Note that in view of the above result the condition $T_{abc}=0$ entails 
$$
\bnb_cP^{ab}=\fr{1}{p}E_cP^{ab},\ \ \bnb_c\Pi^{ab}=\fr{1}{n-p}W_c\Pi^{ab}.
$$
Therefore all the conditions coming from (\ref{def-M}) and (\ref{def-N}) are summarized in $T_{abc}=0$.
The geometric significance of this condition will be investigated in section \ref{sufficient}. 
\label{remark}
\end{remark}

Some of the first integrability conditions achieve a great simplification if $T_{abc}$ vanishes. For instance 
(\ref{cond-ligaduras-down})-(\ref{mixed-invariance}) become zero identically under this condition as is obvious from 
propositions \ref{ftp} and \ref{up-down} so we do not need to care about these integrability
conditions any more. Equation
(\ref{cond-g1}) acquires the invariance law
\be
\lie T^d_{\ cab}=0.
\label{cond-g11}
\ee
Other simplifications will be shown in the forthcoming analysis.

\medskip
\noindent
{\em Equations (\ref{cond-d}) and (\ref{cond-e}).} If $T_{abc}=0$ then $\La^a_{\ bc}=0$, $\bLa^a_{\ bc}=0$ 
so these equations take the form 
\bea
-\bphi_qP^q_{\ r}T^r_{\ ceb}=2\lie\bnb_{[e}L^0_{b]c}+\bchi_rE^r_{\ ceb},\label{cond-df}\\ 
-\bchi_q\Pi^q_{\ r}T^r_{\ ceb}=2\lie\bnb_{[e}L^1_{b]c}+\bphi_rF^r_{\ ceb},\label{cond-ef}  
\eea
where
\bea
E^r_{\ ceb}=\fr{2}{2-p}(\Pi^q_{\ c}\Pi^r_{\ [e}L^0_{\ b]q}+\Pi^r_{\ c}\Pi^q_{\ [e}L^0_{b]q}-\Pi^{rq}\Pi_{c[e}L^0_{b]q})\\
F^r_{\ ceb}=\fr{2}{2-n+p}(P^q_{\ c}P^r_{\ [e}L^1_{\ b]q}+P^r_{\ c}P^q_{\ [e}L^1_{b]q}-P^{rq}P_{c[e}L^1_{b]q}).
\eea
From (\ref{cond-df}) and (\ref{cond-ef}) we find the conditions of complete integrability
$$
P^q_{\ r}T^r_{\ ceb}=0,\ \Pi^q_{\ r}T^r_{\ ceb}=0\Longrightarrow T^q_{\ ceb}=0,
$$
so equation (\ref{cond-g11}) is trivially fulfilled.
To proceed further with the calculations we need a lemma 
\begin{lemma}
If $\bnb_aP^b_{\ c}=0$ then 
$$
L^0_{bq}\Pi^q_{\ c}=0,\ \ L^1_{bq}P^q_{\ c}=0.
$$
\end{lemma}

\begin{proof}
These properties are proven through the Ricci identity applied to the tensors $P^b_{\ c}$, $\Pi^b_{\ c}$ which 
take a remarkably simple form under our conditions (we only perform the calculations for the tensor $P^a_{\ b}$)
\be
0=\bnb_e\bnb_aP^b_{\ c}-\bnb_a\bnb_eP^b_{\ c}=\bR^b_{\ qea}P^q_{\ c}-\bR^q_{\ cea}P^b_{\ q},
\label{ricci-p}
\ee 
whence
$$
L^0_{bq}\Pi^q_{\ c}=2P^d_{\ r}\bR^r_{\ qdb}\Pi^q_{\ c}+\fr{2}{p}(\Pi^d_{\ b}P^r_{\ q}\bR^q_{\ rds}\Pi^s_{\ c}).
$$
The first term of this expression is zero according to (\ref{ricci-p}) and the second one can be transformed 
by means of the first Bianchi identity into
$$
\Pi^d_{\ b}\Pi^s_{\ c}P^r_{\ q}\bR^q_{\ rds}=-\Pi^d_{\ b}\Pi^s_{\ c}(P^r_{\ q}\bR^q_{\ dsr}+P^r_{\ q}\bR^q_{\ srd})=
-\Pi^d_{\ b}\Pi^s_{\ c}(P^q_{\ d}\bR^r_{\ qsr}+P^q_{\ s}\bR^r_{\ qrd}),
$$
which also vanishes.\qed
\end{proof}
Therefore conditions (\ref{cond-df}) and (\ref{cond-ef}) are further simplified to 
\be
\lie\bnb_{[e}L^0_{b]c}=0,\ \ \ 
\lie\bnb_{[e}L^1_{b]c}=0.
\label{further}
\ee
It is our next aim to show that indeed these two equations are identities if $T^d_{\ cab}=0$.
\begin{lemma}
If $p\neq 3$, $n-p\neq 3$ and $\bnb_aP^b_{\ c}=0$ then
$$
T^d_{\ cab}=0\Longrightarrow \bnb_{[e}L^0_{b]c}=0,\ \ \bnb_{[e}L^1_{b]c}=0
$$
\label{long-lema}
\end{lemma}
\begin{proof}
To prove this we start from the identity
\bea
(2-p)[\bnb_e(P^e_{\ q}T^q_{\ cab})+\bnb_b(P^d_{\ q}T^q_{\ cda})-\bnb_a(P^d_{\ q}T^q_{\ cdb})]=2P^e_{\ c}\bnb_{e}L^0_{[ba]}
-2P^d_{\ c}\bnb_{[b|}L^0_{d|a]}+\nonumber\\
+2p\bnb_{[b}L^0_{a]c}+2P^d_{\ [b}\bnb_{a]}L^0_{dc}
-2P^e_{\ [b|}\bnb_eL^0_{|a]c}+2P^{qe}\bnb_eL^0_{[a|q}P_{|b]c}+2P^{qd}P_{c[a}\bnb_{b]}L^0_{dq},\nonumber\\
\label{ident-1} 
\eea
which is easily obtained from equation (\ref{tensor-tt}) and the second Bianchi identity for the tensor 
$\bR^a_{\ bcd}$ if the condition $\bnb_aP^b_{\ c}=0$ holds. In our case the left hand side of this identity vanishes so we only need to study the 
right hand side equated to zero. Contracting such equation with $P^{ca}$ we get
$$
2(1-p)P^{ae}\bnb_eL^0_{ba}-2(1-p)P^{da}\bnb_bL^0_{da}+2P^d_{\ b}P^{ac}\bnb_aL^0_{dc}-2P^e_{\ b}P^{ac}\bnb_eL^0_{ac}=0,
$$
and a further contraction with $P_a^{\ r}$ yields
$$
P^{qe}\bnb_eL^0_{aq}=P^{qd}\bnb_aP_{qd}.
$$
This last property implies that the last two terms of (\ref{ident-1}) are zero and it becomes
\be
2P^e_{\ c}\bnb_{e}L^0_{[ba]}
-2P^d_{\ c}\bnb_{[b|}L^0_{d|a]}
+2p\bnb_{[b}L^0_{a]c}+2P^d_{\ [b}\bnb_{a]}L^0_{dc}=0.
\label{ident-2}
\ee
 If we multiply this last equation by $P_a^{\ r}P_b^{\ s}$  we obtain
\be
S_{cba}-S_{bca}+S_{acb}-S_{cab}+(p-2)(S_{bac}-S_{abc})=0,
\label{Sabc}
\ee
where
$$
S_{abc}\equiv P_a^{\ r}P_b^{\ s}\bnb_rL^0_{sc}.
$$
Permuting indexes in (\ref{Sabc}) we get the equations
\be
\left.
\begin{array}{c}
S_{cba}-S_{bca}+S_{acb}-S_{cab}+(p-2)(S_{bac}-S_{abc})=0\\
S_{abc}-S_{bac}+S_{cab}-S_{acb}+(p-2)(S_{bca}-S_{cba})=0\\
S_{cba}-S_{bca}+S_{bac}-S_{abc}+(p-2)(S_{acb}-S_{cab})=0
\end{array}\right\}.
\ee
Setting the variables $x=S_{cba}-S_{bca}$, $y=S_{acb}-S_{cab}$, $z=S_{bac}-S_{abc}$ we deduce
that previous equations form a homogeneous system in these variables whose matrix is
$$
\left(\begin{array}{ccc}
1&1&p-2\\
2-p&-1&-1\\
1&p-2&1\end{array}\right)\Rightarrow \left|
\begin{array}{ccc}
1&1&p-2\\
2-p&-1&-1\\
1&p-2&1\end{array}\right|=-p(p-3)^2.
$$
So unless $p=3$ ($p=0$ makes no sense in the current context) we conclude that $x=y=z=0$ and hence 
$$
P_{[a}^{\ r}P_{b]}^{\ s}\bnb_rL^0_{sc}=0.
$$
Application of this in the expression resulting of multiplying (\ref{ident-2}) by $P^r_{\ b}$ leads to
$$
P^d_{\ c}\bnb_aL^0_{db}-P^e_{\ c}\bnb_eL^0_{ab}+(p-1)(P^r_{\ b}\bnb_rL^0_{ac}-P^r_{\ b}\bnb_aL^0_{rc})=0.
$$
By setting $Q_{acb}\equiv P^d_{\ c}\bnb_aL^0_{db}-P^e_{\ c}\bnb_eL^0_{ab}$ we can rewrite this as
$$
Q_{acb}-(p-1)Q_{abc}=0,\Rightarrow Q_{abc}-(p-1)Q_{acb}=0,
$$
which entails $Q_{abc}=0$ (recall that $p\neq 1$ by definition of $L^0_{ab}$). 
This last property applied to (\ref{ident-2}) yields 
$$
\bnb_{[b}L^0_{a]c}=0,
$$ 
as desired. The result for $L^1_{ab}$ is proven in a similar way.\qed
\end{proof}

\medskip
\noindent
{\em Equations (\ref{cond-a1}) and (\ref{cond-a2})} The analysis of these conditions is performed 
by means of the following result.
\begin{proposition}
If $T_{abc}=0$ then 
\bnr
\fr{1}{2-p}(L^0_{ab}-L^0_{ba})-\fr{1}{p}(\bnb_aE_b-\bnb_bE_a)=0,\\ 
\fr{1}{2-n+p}(L^1_{ab}-L^1_{ba})-\fr{1}{n-p}(\bnb_aW_b-\bnb_bW_a)=0.
\enr
\end{proposition}
\begin{proof}
We only carry on the proof for the first identity as the calculations are similar for the second one. From 
(\ref{L0}) and applying the first Bianchi identity is easy to obtain
\be
L^0_{ab}-L^0_{ba}=\fr{2(2-p)}{p}P^r_{\ q}\bR^q_{\ rab}.
\label{uno}
\ee
On the other hand
\be
\bnb_aE_b-\bnb_bE_a=P^{qr}(\bnb_a\bnb_bP_{qr}-\bnb_b\bnb_aP_{qr})-\bnb_aP^{qr}\bnb_bP_{qr}+\bnb_bP^{qr}\bnb_aP_{qr}.
\label{qxi}
\ee
If we impose now the condition $T_{abc}=0$ then combination of proposition \ref{ftp} and remark \ref{remark} entails
$$
\bnb_aP^{qr}\bnb_bP_{qr}=-\fr{1}{p}E_aE_b=\bnb_bP^{qr}\bnb_aP_{qr},
$$
so (\ref{qxi}) becomes
\be
\bnb_aE_b-\bnb_bE_a=2P^q_{\ r}\bR^r_{\ qab}.
\label{dos}
\ee
Combination of (\ref{uno}) and (\ref{dos}) leads to the desired result.\qed
\end{proof}

\medskip
\noindent
{\em Equations (\ref{cond-b}) and (\ref{cond-c})}
\begin{proposition}
If $T^d_{\ cab}=0$ and $T_{abc}=0$ then (\ref{cond-b}) and (\ref{cond-c}) are 
identities.
\end{proposition}
\begin{proof}
The left hand side of both equations vanishes trivially if $T^d_{\ cab}=0$ so we just need to show that 
the right hand side vanishes as well. The characterization of the condition $T_{abc}=0$ in terms of the 
covariant derivatives of the projectors entails
$$
\Up^{sc}_{\ b}=\fr{2-p}{p}E_bP^{sc},\ \ \bUp^{sc}_{\ b}=\fr{2-n+p}{n-p}W_b\Pi^{sc}, 
$$
which means that the terms of (\ref{cond-b}) and (\ref{cond-c}) containing these tensors are zero.
The property $\bnb_cP^a_{\ b}=\bnb_c\Pi^a_{\ b}=0$ can be used now to get rid of some terms and simplify others
on these couple of equations getting
\bnr
0=\bchi_a\bnb_{[c}E_{b]}+\bchi_{[b}\bnb_{c]}E_a+\bchi_a\bnb_{[b}E_{c]}+\bchi_{[c}\bnb_{b]}E_a\\
0=\bphi_a\bnb_{[c}W_{b]}+\bphi_{[b}\bnb_{c]}W_a+\bphi_a\bnb_{[b}W_{c]}+\bphi_{[c}\bnb_{b]}W_a,
\enr
which is obviously zero.\qed
\end{proof}

All our calculations are thus summarized in the next result which is one of the most important 
 of this paper.
\begin{theorem}[Complete integrability conditions]
The necessary and sufficient geometric conditions for a space $(V,\G)$ to possess $N$ linearly independent 
bi-conformal vector fields associated to the projectors $P_{ab}$ $\Pi_{ab}$ not projecting on a subspace of 
dimension three are 
\begin{center}
\fbox{$T_{abc}=0,\ \ \ T^d_{\ cab}=0$.}
\end{center}
\label{complete-integrability}
\end{theorem}

  
\section{Geometric characterisation of conformally separable pseudo-Riemannian manifolds}
\label{sufficient}
Once we have found the mathematical characterization of the spaces admitting a maximum number of bi-conformal vector 
fields we must next settle if there is actually any space whose metric tensor complies with the conditions stated in theorem
\ref{complete-integrability} or on the contrary there are no pseudo-Riemannian manifolds fulfilling such requirement.
Indeed we will find that each geometric condition has a separate meaning related with the geometric characterization of 
certain {\em separable} pseudo-Riemannian manifolds. Hence the tensors $T_{abc}$ and $T^a_{\ bcd}$ bear a geometric 
interest on their own regardless to the existence or not of bi-conformal vector fields on the space $(V,\G)$ where they are 
defined. Before addressing this issue 
we need some preliminary definitions.
\begin{definition}
The pseudo-Riemannian manifold $(V,\G)$ is said to be separable at the point $q\in V$ if there exists a 
local coordinate chart 
$x\equiv\{x^1,\dots,x^n\}$ based at $q$ in which the metric tensor takes the form
\be
\rmg_{ab}(x)=\left\{
\begin{array}{c}
\rmg_{\alpha\beta}(x),\ 1\leq\alpha,\beta,\leq p\\
\rmg_{AB}(x),\ p+1\leq A,B\leq n\\
0\ \ \mbox{otherwise.}
\end{array}\right.
\label{metric-separable}
\ee
$(V,\G)$ is separable if it is so at every point $p\in V$. Any of the metric tensors $\rmg_{\alpha\beta}$, $\rmg_{AB}$ 
in which $\rmg_{ab}$ is split shall be called leaf metric of the separation.
\label{separable}
\end{definition}
This same concept is already defined in \cite{YANOMASAHIRO} for the case of Riemannian manifolds (they call these manifolds 
{\em locally product Riemannian manifolds}) 

Henceforth all our results will deal with separable pseudo-Riemannian manifolds but they have a clear extension 
to the case in which only local separability at a point holds. From now on 
when working with separable spaces written in the form of (\ref{metric-separable})
we adopt the convention that 
Greek letters label indexes associated to one of the leaf metrics whereas uppercase Latin characters 
are used for the other one. 
The coordinate system of definition \ref{metric-separable} is fully adapted to the decomposition of the metric
tensor but in general we cannot expect this to be the case when dealing with arbitrary separable pseudo-Riemannian 
manifolds. Therefore it would be desirable to have a result characterizing separable pseudo-Riemannian manifolds 
in a coordinate-free way. 
Nonetheless no general result characterizing separable spaces in a co\-or\-di\-na\-te-free way is known 
although there
are already available results for particular cases of separable spaces.  
Before considering them we need to give a brief account of the most studied types of separable pseudo-Riemannian 
manifolds in the literature.

\begin{definition}
Let $x\equiv\{x^a\}$, $a=1\dots n$ be the local coordinate system introduced in definition \ref{separable}.  
A separable manifold can then be classified in terms of the form the line element $ds^2$ takes in these coordinates as 
\begin{enumerate}
\item decomposable or reducible:\\
 $ds^2=\rmg_{\alpha\beta}(x^{\epsilon})dx^{\alpha}dx^{\beta}+\rmg_{AB}(x^C)dx^Adx^B$, 
\item semi-decomposable, semi-reducible or warped product:\\ 
$ds^2=\rmg_{\alpha\beta}(x^{\epsilon})dx^{\alpha}dx^{\beta}+\Xi(x^{\epsilon})G_{AB}(x^C)dx^Adx^B$, $\Xi(x^{\epsilon})$ warping
factor,
\item generalized decomposable or double warped:\\ 
$ds^2=\Xi_1(x^{\C})G_{\alpha\beta}(x^{\epsilon})dx^{\alpha}dx^{\beta}+\Xi_2(x^{\epsilon})G_{AB}(x^C)dx^Adx^B$,\ \-\ \-
$\Xi_1(x^C)$, $\Xi_2(x^{\epsilon})$ warping factors,
\item conformally reducible:\\
$ds^2=\Xi(x^a)(G_{\alpha\beta}(x^{\epsilon})dx^{\alpha}dx^{\beta}+G_{AB}(x^C)dx^Adx^B)$,
\item conformally separable or double twisted:\\
$ds^2=\Xi_1(x^a)G_{\alpha\beta}(x^{\epsilon})dx^{\alpha}dx^{\beta}+\Xi_2(x^a)G_{AB}(x^C)dx^Adx^B$.
\item bi-conformally flat:\\
$ds^2=\Xi_1(x^a)\eta_{\alpha\beta}(x^\epsilon)dx^{\alpha}dx^{\beta}+\Xi_2(x^a)\eta_{AB}(x^C)dx^Adx^B$,
 $\eta_{\alpha\beta}$, $\eta_{AB}$ flat metrics of dimension $p$ and $n-p$ respectively.
\end{enumerate}
\label{classification}
\end{definition}
 
There are here classic results in Differential Geometry characterizing invariantly some of the above 
types of separable spaces. 

\begin{theorem}
A pseudo-Riemannian manifold is decomposable if and only if either of the conditions 
written below holds
\begin{enumerate}
\item  The manifold possesses two orthogonal families of foliations by totally geodesic hypersurfaces. 
\item There exists a symmetric and idempotent tensor $P_{ab}\neq\rmg_{ab}$ such that $\nb_aP_{bc}=0$. 
In this case the tensor $P_{ab}$ is one of the leaf metrics 
of the decomposition being $\rmg_{ab}-P_{ab}$ the other one.
\end{enumerate}
\label{sep-theorem}
\end{theorem}

\begin{proof} The original proof of the first point was given in \cite{BOMPIANI} 
(see also p.186 of \cite{EISENHART} and p. 420 of \cite{YANOMASAHIRO}). The second point 
was proven in \cite{PETROV} in the context of General Relativity
(\cite{YANOMASAHIRO} also proves this result indirectly)
but we provide its proof as it employs techniques needed later. 
Choose an orthonormal co-basis 
$\{\bz^1,\dots,\bz^n\}$ adapted to $P_{ab}$, that is to say, its $p$ first elements are dual to a basis of 
the eigenspace of $P^a_{\ b}$ with nonvanishing eigenvalue (observe that under the assumptions of this theorem $P_{ab}$ 
is an orthogonal projector). In terms of this basis $P_{ab}$ takes the form (we use index-free notation and index label
splitting as in definition \ref{separable})
$$
\bP=\sum_{\alpha=1}^p\epsilon_{\alpha}\bz^{\alpha}\otimes\bz^{\alpha},
$$
where $\epsilon_{\alpha}=\pm 1$ (the exact value for each index $\alpha$ will depend on the signature of $\rmg_{ab}$).
Now since 
$$
\nb_c\bz^{\alpha}=-\g^{\alpha}_{\ rc}\bz^r=-\g^{\alpha}_{\ \beta c}\bz^{\beta}-\g^{\alpha}_{\ Bc}\bz^B, 
$$
we have
$$
0=\nb_c\bP=-\sum_{\alpha=1}^{p}\epsilon_{\alpha}[\g^{\alpha}_{\ \beta c}(\bz^{\beta}\otimes\bz^{\alpha}+
\bz^{\alpha}\otimes\bz^{\beta})+
\g^{\alpha}_{\ Bc}(\bz^{B}\otimes\bz^{\alpha}+\bz^{\alpha}\otimes\bz^B)],
$$
which entails $\g^{\alpha}_{\ Bc}=0$. Similarly using the property $\nb_c(\rmg_{ab}-P_{ab})=0$ we can show that 
$\g^B_{\ \alpha c}=0$. These two conditions upon the connection coefficients imply by means of Frobenius theorem that
the distributions spanned by $\{\bz^1,\dots,\bz^p\}$ and $\{\bz^{p+1},\dots,\bz^n\}$ are both integrable. 
In the local coordinate system $\{x^1,\dots,x^n\}$ adapted to the manifolds generated by these distributions 
(i.e. $x^{\alpha}=c^{\alpha}$, $x^A=c^A$) the line element takes the form 
$$
ds^2=\rmg_{\alpha\beta}(x^a)dx^{\alpha}dx^{\beta}+\rmg_{AB}(x^a)dx^Adx^B.
$$
Moreover if we impose the conditions $\nb_cP_{ab}=\nb_c(\rmg_{ab}-P_{ab})=0$ on this metric 
we get 
$$
\d_A\rmg_{\alpha\beta}=0,\ \ \d_{\alpha}\rmg_{AB}=0,
$$
which implies that the metric is decomposable.
The converse is straightforward if we write the metric in the form of point 1 of definition \ref{classification} 
and realize that the tensor 
$$
P_{ab}=\delta_a^{\alpha}\delta_b^{\beta}\rmg_{\alpha\beta}
$$
has the required properties. \qed
\end{proof}

There are also special characterizations of four dimensional decomposable pseudo-Riemannian manifolds devised 
in the framework of General Relativity using {\em Holonomy groups} (see \cite{HALL}). 

The invariant characterization of conformally separable pseudo-Riemannian manifolds is covered 
by the next result (all the tensors appearing herein are understood in terms of their homonym counterparts of
section \ref{first-integrability}).
\begin{theorem}
A pseudo-Riemannian manifold $(V,\G)$ is conformally separable if and only if 
any of the following conditions are satisfied.
\begin{enumerate}
\item The manifold admits 
two orthogonal families of foliations by totally umbilical hypersurfaces.
The family of first fundamental forms of each hypersurface gives rise to 
the leaf metrics of the decomposition of $\rmg_{ab}$ in the obvious way. 
\item There exists an orthogonal projector 
$P_{ab}$ such that the tensor $T_{abc}$ formed with $P_{ab}$ and its complementary $\Pi_{ab}=\rmg_{ab}-P_{ab}$ is 
zero identically. In such case $P_{ab}$ and $\Pi_{ab}$ are the leaf metrics of the separation.
\end{enumerate}
\label{conf-separable}
\end{theorem}

\begin{proof}
The first point was proven in \cite{YANO2} and the second point is a new result which comes naturally from
the geometric tools of previous sections (only the latter result is shown in this proof).
The proof of the only if implication is theorem 7.2 of \cite{BI-CONFORMAL} 
but for the sake of completeness we 
reproduce its main details here.  Choose the coordinate system and notation of de\-fi\-nition 
\ref{separable} and to keep generality 
assume that the manifold is only separable. In this coordinate system the leaf metrics $P_{ab}$ and $\Pi_{ab}$ look like
$$
P_{ab}=\rmg_{\alpha\beta}\delta^{\alpha}_{\ a}\delta^{\beta}_{\ b},\ \ \ 
\Pi_{ab}=\rmg_{AB}\delta^A_{\ a}\delta^B_{\ b},
$$  
and the non-zero components of the Christoffel symbols are
\bnr
\Chr^{\alpha}_{\ \beta\g}&=&
\fr{1}{2}\rmg^{\alpha\rho}(\d_{\beta}\rmg_{\g\rho}+\d_{\g}\rmg_{\rho\beta}-\d_{\rho}\rmg_{\beta\g}),\ 
\Chr^{\alpha}_{\ \beta A}=\fr{1}{2}\rmg^{\alpha\rho}\d_A\rmg_{\beta\rho},\\
\ \Chr^\alpha_{\ BA}&=&-\fr{1}{2}\rmg^{\alpha\rho}\d_{\rho}\rmg_{BA},
\Chr^A_{\ B\alpha}=\fr{1}{2}\rmg^{AD}\d_{\alpha}\rmg_{BD},\\ 
\Chr^A_{\ \alpha\beta}&=&-\fr{1}{2}\rmg^{AD}\d_D\rmg_{\beta\alpha},\ 
\Chr^A_{\ BC}=\fr{1}{2}\rmg^{AD}(\d_B\rmg_{CD}+\d_C\rmg_{DB}-\d_D\rmg_{BC}),
\enr
where
$$
\rmg^{\alpha\rho}\rmg_{\rho\beta}=\delta^{\alpha}_{\ \rho},\ \ 
\rmg^{AD}\rmg_{DB}=\delta^A_{\ B}.
$$
The only nonvanishing components of $M_{abc}$, $E_a$, $W_a$ are thus
\bea
M_{\alpha AB}=\d_{\alpha}\rmg_{AB},\ \ 
M_{A\alpha\beta}=-\d_{A}\rmg_{\alpha\beta},\nonumber\\
E_{A}=-\d_{A}\log|\mbox{det}(\rmg_{\alpha\beta})|,
W_{\alpha}=-\d_{\alpha}\log|\mbox{det}(\rmg_{AB})|,
\label{M-W}
\eea
from which we deduce that those of $T_{abc}$ are
\be
T_{\alpha AB}=\d_{\alpha}\rmg_{AB}+\fr{1}{n-p}\rmg_{AB}W_{\alpha},\ \ 
T_{A\alpha\beta}=\d_A\rmg_{\alpha\beta}+\fr{1}{p}\rmg_{\alpha\beta}E_A,
\label{couple}
\ee
which are identically zero if the metric tensor represents a conformally separable space.
To prove the converse we use the equivalent condition in terms of the covariant derivative with respect 
to the bi-conformal connection (proposition \ref{ftp})
$$
\bnb_cP_{ab}=-\fr{1}{p}E_cP_{ab},\ \ \bnb_c\Pi_{ab}=-\fr{1}{n-p}W_c\Pi_{ab}.
$$
Proceeding along the same lines as in theorem theorem \ref{sep-theorem} we can easily show that this condition entails that 
the space is separable (we only need to replace the components of the metric connection by those of 
the bi-conformal connection $\bar{\g}^a_{\ bc}$). Thus we are left with the couple of equations
 (\ref{couple}) equalled zero. The general solution of the resulting PDE system is 
$$
\rmg_{\alpha\beta}=G_{\alpha\beta}(x^{\delta})e^{\Lambda_1(x^a)},\ 
\rmg_{AB}=G_{AB}(x^D)e^{\Lambda_2(x^a)},
$$
where $G_{\alpha\beta}$, $G_{AB}$, $\Lambda_1$, $\Lambda_2$ 
are arbitrary functions of their respective arguments 
 with no restrictions other than det$(G_{\alpha\beta})\neq 0$, det$(G_{AB})\neq 0$. Comparing these 
expression with the fifth point of definition \ref{classification} the result follows.\qed
\end{proof}

This theorem clearly states the geometric relevance of $T_{abc}$ as a tool to characterize conformally separable 
pseudo-Riemannian manifolds. In fact 
the condition $T_{abc}=0$ can be re-written in terms of the factors $\Xi_1$ and $\Xi_2$ introduced in the definition
of a conformally separable metric. To that end we use the equivalent condition (\ref{25}) and replace the 1-forms 
$E_a$, $W_a$ by the values given by equation (\ref{M-W}) which are
$$
E_a=-p\Pi_a^{\ r}\d_r\log|\Xi_1|,\ \ W_a=(p-n)P_a^{\ r}\d_r\log|\Xi_2|,
$$
whence
\be
\nb_bP_{ac}=P_{bc}u_a+P_{ab}u_c-P_a^{\ r}u_r\rmg_{bc}-P_c^{\ r}u_r\rmg_{ab},
\label{Xi12}
\ee
where 
$$
u_a=\fr{E_a}{2p}+\fr{W_a}{2(n-p)}.
$$
If $\Xi_1=\Xi_2=\Xi$ then the factor $\Xi$ is invariantly determined by the relation
$$
\fr{E_a}{p}+\fr{W_a}{n-p}=-\d_a\log|\Xi|,
$$ 
which means that the 1-form $u_a$ of (\ref{Xi12}) is exact. This characterization of conformally reducible 
pseudo-Riemannian manifolds was already proven in \cite{ANCIKOV} by other means. Finally let us point out that
studies of conformally reducible spaces in the particular case of dimension four have been performed in \cite{CAROT}.

Bi-conformally flat spaces are a particular and interesting case of conformally se\-pa\-rable pseudo-Riemannian manifolds
so they comply with the condition written above. In this case we can even go further and refine the characterization
in terms of the tensor $T_{abc}$ for these spaces.

\begin{theorem}
A conformally separable pseudo-Riemannian manifold with leaf metrics of rank greater than 3 
is bi-con\-for\-ma\-lly flat if and only if the tensor
 $T^a_{\ bcd}$ constructed from its leaf metrics is identically zero.
\label{bi-conformal-char} 
\end{theorem}

\begin{proof} 
We choose the same coordinates and notation for our conformally separable space
 as in example \ref{example1}.
Using the components of the bi-conformal connection calculated there for this metric 
we deduce after some algebra
\bea
\bR^{\alpha}_{\ \beta\phi\g}&=&R^{\alpha}_{\ \beta\phi\g},\ \bR^{A}_{\ BCD}=R^A_{\ BCD},\ 
\bR^{\alpha}_{\ \beta F\phi}=\d_F\bar{\Chr}^{\alpha}_{\ \phi\beta},\ \bR^A_{\ BF\phi}=-\d_{\phi}\bar{\Chr}^A_{\ FB}
\nonumber\\
L^0_{\alpha\beta}&=&2R_{\alpha\beta}+\fr{R^{\g}_{\ \g}}{1-p}\rmg_{\alpha\beta},\ \ 
L^1_{AB}=2R_{AB}+\fr{R^C_{\ C}}{1-n+p}\rmg_{AB},\ L^0_{\alpha A}=L^1_{A\alpha}=0,\nonumber\\
L^0_{\ A\alpha}&=&\fr{2(2-p)}{p}\d_A\bar{\Chr}^{\rho}_{\ \alpha\rho},\ L^1_{\alpha A}=\fr{2(2-n+p)}{n-p}\d_{\alpha}\bar{\Chr}^{R}_{\ AR}
\label{cua} 
\eea
where $R_{\alpha\beta}$, $R_{AB}$, $R^{\g}_{\ \g}$ and $R^{C}_{\ C}$ are the Ricci tensors and Ricci scalars of 
$R^{\alpha}_{\ \beta\phi\g}$ and $R^A_{\ BCD}$ respectively (observe that 
these curvature tensors are calculated from the leaf metrics and not 
from their conformal counterparts $G_{\alpha\beta}$ and $G_{AB}$). 
The tensor components not shown in (\ref{cua}) are equal
 to zero. From this and equations (\ref{tensor-tt}), (\ref{new-components}) we get 
that the only nonvanishing components of the tensor $T^a_{\ bcd}$ are
\be
T^{\alpha}_{\ \beta\phi\g}=2C^{\alpha}_{\ \beta\phi\g},\ \ \ T^A_{\ BCD}=2C^A_{\ BCD},
\label{weyl}
\ee
being $C^{\alpha}_{\ \beta\phi\g}$ and $C^A_{\ BCD}$ the Weyl tensors constructed from each leaf metric 
through the relations
\bnr
C^{\alpha}_{\ \beta\phi\g}=R^{\alpha}_{\ \beta\phi\g}+\fr{1}{2-p}(\rmg_{\beta[\g}L^0_{\phi]\rho}\rmg^{\rho\alpha}
+\delta^{\alpha}_{\ [\phi}L^0_{\g]\beta}),\\ 
C^A_{\ BCD}=R^A_{\ BCD}+\fr{1}{2-n+p}(\rmg_{B[D}L^1_{C]E}\rmg^{EA}+\delta^A_{\ [C}L^1_{D]B}).
\enr
Hence $T^{\alpha}_{\ \beta\phi\g}$ and  $T^A_{\ BCD}$ are both zero if and only if the leaf metrics $\rmg_{\alpha\beta}$
and $\rmg_{AB}$ are both conformally flat
which proves the theorem.\qed 
\end{proof}

\begin{remark}
Note that the tensors $\bR^{\alpha}_{\ \alpha F\phi}$ and $\bR^{A}_{\ AF\phi}$ do not vanish in general 
which means that the bi-conformal connection does not stem from a metric tensor in this case.
\end{remark}
From the calculations performed above is clear that theorem \ref{bi-conformal-char} can be generalized to 
conformally separable spaces in which only one of the leaf metrics is conformally flat.

\begin{theorem}
Under the assumptions of theorem \ref{bi-conformal-char} a leaf metric is conformally flat if and only if 
the tensor $\ ^{||}C^a_{\ bcd}$ calculated from the leaf metric $P_{ab}$ is equal to zero.  
\label{bi-conformal-char-2}
\end{theorem}
\begin{proof}
From the proof of theorem \ref{bi-conformal-char} we deduce that for a pseudo-Riemannian conformally separable 
manifold the only non-vanishing components of $\ ^{||}C^a_{\ bcd}$ are 
$$
\ ^{||}C^{\alpha}_{\ \beta\g\phi}=T^{\alpha}_{\ \beta\g\phi}=2C^{\alpha}_{\ \beta\g\phi},\ \ 
$$
so the vanishing of $\ ^{||}C^a_{\ bcd}$ implies that the Weyl tensor calculated from the corresponding leaf metric is zero 
as well.\qed
\end{proof}

\begin{remark}
This last theorem can be restated as the converse of theorem \ref{ppC}.
\end{remark}

In the case of any of the leaf metrics being of rank three is clear from the above that the corresponding tensor 
$\ ^{||}C^a_{\ bcd}$ will be zero as the Weyl tensor of any three dimensional 
pseudo-Riemannian metric 
vanishes identically. Hence the results presented so far cannot be used to characterize 
conformally separable pseudo-Riemannian manifolds with conformally flat leaf metrics.
 This lacking is remedied in the next theorem.
\begin{theorem}
A conformally separable pseudo-Riemannian manifold has a conformally flat leaf metric of rank three if and only
if the condition
\be
\bnb_{[a}L^0_{b]c}=0, 
\label{rank-3}
\ee
holds for the leaf metric tensor $P_{ab}$.  
\label{case-3}
\end{theorem}   
\begin{proof}
To show this we will rely on the notation and calculations performed in the proof of theorem 
\ref{bi-conformal-char}. 
If the manifold is conformally separable then the only non-zero components of the tensor $\bnb_aL^0_{bc}$ are
\bnr
\bnb_{\alpha}L^0_{\beta\epsilon}&=&\nb_{\alpha}L^0_{\beta\epsilon},\ \ 
\bnb_{\alpha}L^0_{B\epsilon}=\fr{2(2-p)}{p}(\d^2_{\alpha B}\bar{\Chr}^{\rho}_{\ \epsilon\rho}-
\bar{\Chr}^{\delta}_{\ \alpha\epsilon}\d_B\bar{\Chr}^{\rho}_{\delta\rho})
,\ \bnb_AL^0_{\beta\epsilon}=\d_AL^0_{\beta\epsilon},\\
\bnb_AL^0_{B\epsilon}&=&\d_AL^0_{B\epsilon}-\bar{\Chr}^C_{AB}L^0_{C\epsilon}
\enr
where $\nb_{\alpha}$ is the covariant derivative compatible with $\rmg_{\alpha\beta}$. Trivially
\be
\bnb_{[\alpha}L^0_{\beta]\epsilon}=\nb_{[\alpha}L^0_{\beta]\epsilon},\ \
\label{eq}
\ee
here the tensor $\nb_{[\alpha}L^0_{\beta]\epsilon}$ is the Cotton-York tensor of the 3-metric $\rmg_{\alpha\beta}$ 
which vanish if and only if such metric is conformally flat. Therefore to finish the proof of this theorem 
we must show that all the remaining components of $\bnb_{[a}L^0_{b]c}$ are zero for any conformally separable 
metric $\rmg_{ab}$. These are
$$
\bnb_{[A}L^0_{B]\epsilon}=\fr{2(2-p)}{p}\d^2_{[AB]}\bar{\Chr}^{\rho}_{\ \epsilon\rho},\ 
2\bnb_{[\alpha}L^0_{B]\epsilon}=\fr{2(2-p)}{p}(\d^2_{\alpha B}\bar{\Chr}^{\rho}_{\ \epsilon\rho}-
\bar{\Chr}^{\delta}_{\ \alpha\epsilon}\d_B\bar{\Chr}^{\rho}_{\delta\rho})-\d_BL^0_{\alpha\epsilon}.
$$
Clearly the first expression is zero and the second one is worked out by replacing 
the connection coefficients by their expressions given in (\ref{new-components}) and 
using of the identity 
\bnr
L^0_{\alpha\beta}&=&L^0_{\alpha\beta}(G)+(2-p)(2\sigma_{\alpha\beta}+G_{\alpha\beta}(\d\sigma)^2),\\ 
\sigma&=&\fr{1}{2}\log|\Xi_1(x^a)|,\ \sigma_{\alpha\beta}=\hat{\nb}_{\alpha}\hat{\nb}_{\beta}\sigma
-\d_{\alpha}\sigma\d_{\beta}\sigma,\ 
(\d\sigma)^2=G^{\alpha\beta}\d_{\alpha}\sigma\d_{\beta}\sigma,
\enr
where $L^0_{\alpha\beta}(G)$ is calculated using curvature tensors computed from $G_{\alpha\beta}$ and 
$\hat{\nb}$ is the connection compatible with this metric. The sought result comes after some 
simple algebraic manipulations.
\qed
\end{proof}

Theorems \ref{bi-conformal-char}, \ref{bi-conformal-char-2} and \ref{case-3}
 supply an invariant geometric 
characterization of conformally separable pseudo-Riemannian manifolds with 
conformally flat leaf metrics. This means that the conditions imposed 
by theorem \ref{complete-integrability} are satisfied by a nontrivial set of pseudo-Riemannian manifolds, namely, 
that formed by bi-conformally flat spaces. These spaces can thus be characterized as those admitting a maximum number 
of bi-conformal vector fields. On the other hand it is straightforward to check (proposition 6.1 of \cite{BI-CONFORMAL}) 
that for these spaces 
any conformal Killing vector of the leaf metrics is a bi-conformal vector field of the whole metric $\rmg_{ab}$. As the number 
of conformal Killing vectors for each leaf metric is the biggest possible as well we get at once that for any 
bi-conformally flat space the total number of linearly independent bi-conformal vector fields is 
$$
N=\fr{1}{2}p(p+1)+\fr{1}{2}(n-p)(n-p+1),\ \ p,\ n-p\neq 2
$$       
so the upper bound placed by theorem \ref{bounded} is actually achieved. Summing up we obtain the following 
result
\begin{theorem}
A pseudo-Riemannian manifold possesses $N$ bi-conformal vector fields ($P^a_{\ a}\neq 3$, $\Pi^a_{\ a}\neq n-3$) 
if and only if it is bi-conformally flat.\qed
\label{N}
\end{theorem}

Bi-conformally flat spaces in which any of the leaf metrics has rank three still admit $N$ linearly independent bi-conformal 
vector fields (in fact the complete integrability conditions are also satisfied for these spaces as a result 
of theorem \ref{case-3}). However, we do not know yet if there are spaces with $N$ linearly independent 
bi-conformal vector fields with either of the projectors $P_{ab}$ or $\Pi_{ab}$ projecting on a 3-dimensional vector 
space other than bi-conformally flat spaces. This is so because in such case the complete integrability conditions 
(\ref{further}) may in principle be fulfilled by other conformally separable spaces not necessarily with conformally 
flat leaf metrics. The true extent of these assertion and the complete characterization of spaces with $N$ linearly 
independent bi-conformal vector fields under these circumstances will be placed elsewhere. 

All in all the geometric conditions proven in theorems 
\ref{conf-separable}, \ref{bi-conformal-char}, \ref{bi-conformal-char-2} and \ref{case-3} provide 
a set of equations which can be used to search systematically for conformally flat foliations of a given 
pseudo-Riemannian manifold $(V,\G)$. For if we set to $P_{ab}$ the leaf metric of such foliation
as we did before then the differential equations 
\be
T_{abc}=0\ \ \ ^{||}C^a_{\ bcd}=0,
\label{cfpg}
\ee  
could in principle allow us to find $P_{ab}$ in a local coordinate system and hence the foliation. 
Of course the second condition can be replaced by $\bnb_{[a}L^0_{b]c}$ if we look for leaf metrics of rank three or 
$T^a_{\ bcd}=0$ if we wish to check whether the metric is bi-conformally flat or not. If such 
system does not admit any solution then previous results guarantee that our metric cannot be foliated by conformally
flat hypersurfaces whose first fundamental forms are leaf metrics of a conformal separation. 

A natural question is now the generalization of the above conditions to pseudo-Riemannian manifolds which are not 
conformally separable, i.e. is there a set of equations similar to (\ref{cfpg}) for cases with $T_{abc}\neq 0$?
A first thought could be that (\ref{cfpg}) or its generalizations for three dimensional cases with the condition 
$T_{abc}=0$ removed would still be true for metrics not conformally separable. The results of the calculations 
performed in example \ref{example2} hint towards this direction.

\section{Examples}
\label{examples}
\begin{example}
As our first example we consider the four dimensional pseudo-Rie\-man\-ni\-an manifold with metric given by
$$
ds^2=(\Psi^2\sin^2\z-\alpha^2)dt^2+2\Psi^2\sin^2\z d\phi dt+B^2(dr^2+r^2d\z^2)+\Phi^2d\phi^2,
$$
where the coordinate ranges are $-\infty<t<\infty,\ 0<r<\infty,\ 0<\z<\pi,\ 0<\z<2\pi$ and
 the functions $\Psi$, $\alpha$, $B$ and $\Phi$ only depend on the the coordinates $r$, $\z$. 
We will try to find out the conditions under which the metric 
is conformally separable with the hypersurfaces $t=cons$ as one of the leaf metrics.  
A simple calculation shows that 
the projector $P^a_{\ b}$ coming from these hypersurfaces is (now and henceforth all the components omitted in 
an explicit tensor representation are understood to be zero)
$$
P^r_{\ r}=P^{\z}_{\ \z}=P^{\phi}_{\ \phi}=1,\ P^{\phi}_{\ t}=\fr{\Psi^2}{\Phi^2},
$$
which entails
$$
P_{tt}=\fr{\Psi^4}{\Phi^2}\sin^2\z,\ P_{rr}=B^2,\ P_{\z\z}=r^2B^2,\ P_{\phi\phi}=\Phi^2\sin^2\z,\ 
P_{t\phi}=\Psi^2\sin^2\z.
$$
From here we can calculate the components of the tensor $T_{abc}$ and set them equal to zero. After doing that we find 
the following independent conditions
\bnr
& &-\Psi\Phi_r+\Psi_r\Phi=0,\ \ -\Psi\Phi_{\z}+\Psi_{\z}\Phi=0,\\
& &(\alpha^2\Phi^2+\Psi^2(2\Psi^2-\Phi^2)\sin^2\z)(-\alpha\Phi^3\alpha_r+\Psi\sin^2\z(\Psi^3\Phi_r+\Phi\Psi_r(\Phi^2-2\Psi^2)))=0,\\
& &\Psi^4\Phi_{\z}\sin^2\z-\Phi\Psi^3(\Psi\cos\z+2\Psi_{\z}\sin\z)\sin\z+\Phi^3(-\alpha\alpha_z+\Psi\sin\z(\Psi\cos\z+\Psi_{\z}\sin\z))=0,
\enr
which are fulfilled if and only if
$$
\Phi=\Psi=k\sin\z,\ \ \alpha=\pm k\sin^2\z,\ k=\mbox{cons.}
$$
Under these conditions the metric takes the form
$$
ds^2=k^2\sin^4\z(2dtd\phi+d\phi^2)+B^2(dr^2+r^2d\z^2),
$$
which is conformally separable as desired. 
\end{example}
\begin{example}
In the foregoing results we have only concentrated on conformally separable 
pseudo-Riemannian manifolds but nothing was said about manifolds with conformally flat slices and not 
conformally separable. To illustrate this case let us consider the four dimensional pseudo-Riemannian manifold 
given in local coordinates $\{x^1,x^2,x^3,x^4\}$ by
\be
ds^2=\Phi(x)[(dx^1)^2+(dx^2)^2+(dx^3)^2]+2\sum_{i=1}^3\beta_i(x)dx^idx^4+\Psi(x)(dx^4)^2,
\label{general-foliation}
\ee
where $x=\{x^1,x^2,x^3,x^4\}$ and $\Phi(x)$, $\beta_i(x)$, $\Psi(x)$ are functions at least $C^3$ in an open domain.
Clearly the above line element is the most general four dimensional metric admitting a conformally flat foliation 
by three dimensional Riemannian hypersurfaces. 
The nonzero components of the orthogonal projector $P^{a}_{\ b}$ associated to this foliation are
\be
P^1_{\ 1}=P^2_{\ 2}=P^3_{\ 3}=1,\ P^i_{\ 4}=\beta_i(x)/\Phi(x),\ i=1,2,3,
\label{projection}
\ee
from which we easily get
$$
P_{11}=P_{22}=P_{33}=\Phi(x),\ P_{i4}=\beta_i(x),\ i=1,2,3,\ P_{44}=\sum_{i=1}^3\beta^2_i(x)/\Phi(x).
$$
Using this we can check condition (\ref{rank-3}) and see what is obtained. 
This is a rather long calculation which is easily 
performed with any of the computer algebra systems available today (the system used here was GRTensorII).
The result is that the tensor $\bnb_{[a}L^0_{b]c}$ does not vanish in this case although
a calculation using (\ref{projection}) shows the important property
\be
P^r_{\ a}P^s_{\ b}P^q_{\ c}\bnb_{[r}L^0_{s]q}=0.
\label{changed-condition}
\ee
\begin{theorem}
A necessary condition that a four dimensional pseudo-Riemannian manifolds can be foliated by 
conformally flat Riemannian hypersurfaces with associated orthogonal projector $P_{ab}$ is equation 
(\ref{changed-condition}).
\end{theorem} 
 This result suggests 
that it may well be possible to generalize the conditions of theorem \ref{case-3} to metrics of arbitrary 
dimension which are not conformally separable replacing these conditions by (\ref{changed-condition}). 
This example can be generalized if we consider pseudo-Riemannian manifolds of higher dimension foliated by hypersurfaces 
of dimension arbitrary and not necessarily Riemannian.
The condition which must be checked in this case is $\ ^{||}C^a_{\ bcd}=0$ being this condition found to be true in 
all the examples tried. Therefore it seems that the
conditions of theorems \ref{bi-conformal-char-2} and \ref{case-3} hold even though the pseudo-Riemannian manifold is not 
conformally separable. We thus feel inclined to believe that these theorems are true 
for any pseudo-Riemannian manifold admitting conformally flat foliations and they can indeed be used to search systematically 
for such foliations.  The true extent of this assertion is under current research. 
\label{example2}
\end{example}

\begin{acknowledgements}
We wish to thank Jos\'e M. M. Senovilla for a careful reading of the manuscript and his many suggested improvements and 
Graham. S. Hall for valuable correspondence.
Financial support from projects  9/UPV00172.310-14456/2002 and BFM 2000-0018 is also gratefully acknowledged.
\end{acknowledgements}

\appendix
\setcounter{section}{0}
\section{Basic identities involving the Lie derivative}
In this appendix we recall some properties of the Lie derivative needed in the main text. 
Despite their basic character, they are hardly presented in basic Differential Geometry 
textbooks and the author is only aware of \cite{YANO,SCHOUTEN} as the only references in 
which they are studied.  
\begin{proposition}
For any symmetric connection $\bnb$ defined in a differentiable manifold $V$, any vector field 
$\xiv$ at least $C^2$ and a tensor field $T^{a_1\dots a_p}_{\ b_1\dots b_q}\in T^p_{q}(V)$ 
we have the following identities
\bea
\lie\bar{\g}^a_{bc}=\bnb_b\bnb_c\xi^a+\xi^d\bar{R}^a_{\ cdb},\label{lie-xi}\\
\hspace{-1cm}\bnb_c\lie T^{a_1\dots a_s}_{\ b_1\dots b_q}-
\lie\bnb_cT^{a_1\dots a_s}_{\ b_1\dots b_q}=\nonumber\\
=-\sum_{j=1}^s(\lie\bar{\g}^{a_j}_{cr})T^{\dots a_{j-1}ra_{j+1}\dots}_{\ b_1\dots b_q}+
\sum_{j=1}^q(\lie\bar{\g}^r_{cb_j})T^{a_1\dots a_s}_{\dots b_{j-1}rb_{j+1}\dots},
\label{lie-conmmutation}\\
\lie \bar{R}^d_{\ cab}=\bnb_a(\lie\bar{\g}^d_{\ bc})-\bnb_b(\lie\bar{\g}^d_{ac})
\label{lie-curvature},
\eea
where $\bar{\g}^a_{\ bc}$ are the components of the connection $\bnb$ and $\bar{R}^a_{\ bcd}$ its
curvature. Furthermore if a metric tensor $\rmg_{ab}$ is set in $V$ and $\nb$ 
is now the metric connection associated to it then
\be
\lie\g^a_{bc}=\fr{1}{2}\rmg^{ae}\left[\nabla_b(\lie\rmg_{ce})+
\nabla_c(\lie\rmg_{be})-\nabla_e(\lie\rmg_{bc})\right].\label{lie-connection}
\ee
\end{proposition}

\section{Finite transformation rules}
In this appendix we show explicitly that the invariance laws found in theorems \ref{six-parameter} 
and \ref{geometric-invariants} for one-parameter groups of bi-conformal transformations do indeed hold for any 
bi-conformal transformation. Let $\rmg_{ab}$ be a pseudo-Riemannian metric and $P_{ab}$, $\Pi_{ab}$ orthogonal and complementary 
projectors with respect to $\rmg_{ab}$. Define now the metric $\tilde{\rmg}_{ab}$ by
$$
\tilde{\rmg}_{ab}=ZP_{ab}+X\Pi_{ab},\ \ \tilde{\rmg}^{ab}=\fr{1}{Z}P_{ab}+\fr{1}{X}\Pi_{ab},\ \tilde{\rmg}_{ap}\tilde{\rmg}^{pb}=
\dd_a^{\ b},
$$
where $Z$ and $X$ are strictly positive $C^2$ functions on the manifold. We are going now to calculate the tensors 
$\bnb_aP^b_{\ c}$, $\La^b_{\ bc}$, $\ ^{||}C^a_{\ bcd}$ using the metric $\tilde{\rmg}_{ab}$ and the projectors $ZP_{ab}$, 
$X\Pi_{ab}$ instead of $\rmg_{ab}$, $P_{ab}$, $\Pi_{ab}$. The 
result will be that these tensors are kept invariant as expected. Note that in principle the metric $\tilde{\rmg}_{ab}$ does not
need to be the bi-conformal transformed of $\rmg_{ab}$ (actually our manifold may even not admit such transformations). We will 
add a prime over the tensor objects calculated with $\tilde{\rmg}_{ab}$ and try to relate these objects to their unprimed 
counterparts. The basic equation is the relation between the primed and unprimed bi-conformal connections
\be
'\bar{\g}^a_{\ bc}-\bar{\g}^a_{\ bc}=\fr{1}{2}(\bar{Z}_bP^a_{\ c}+\bar{Z}_cP^a_{\ b}-\bar{Z}^aP_{bc})+
\fr{1}{2}(\bar{X}_b\Pi^a_{\ c}+\bar{X}_c\Pi^a_{\ b}-\bar{X}^a\Pi_{bc}),
\ee
where $\bar{Z}_b\equiv P_{a}^{\ r}\d_rZ/Z$, $\bar{X}_b\equiv\Pi_{a}^{\ r}\d_rX/X$. Equation (\ref{barcon-bar}) leads
easily to 
$$
'\bnb_aP^b_{\ c}=\bnb_aP^b_{\ c},\ \ \fr{1}{Z}P^{ar}\ '\bnb_r(ZP_{bc})=P^{ar}\bnb_rP_{bc}, 
$$
so $\bnb_aP^b_{\ c}$ and $\La^a_{\ bc}$ are kept invariant. To show the invariance of $\ ^{||}C^a_{\ bcd}$ takes 
more efforts as we need to find the relation between the curvature tensors $'\bR^a_{\ bcd}$ and $\bR^a_{\ bcd}$.     
This is done by means of (\ref{curvature-connection}) getting
\bnr
& &'\bR^a_{\ bcd}=\bR^a_{\ bcd}+\bnb_{[c}\bar{Z}_{d]}P^a_{\ b}+P^a_{\ [d}\bnb_{c]}\bar{Z}_b-P_{b[d}\bnb_{c]}\bar{Z}^a+
\bnb_{[c}\bar{X}_{d]}\Pi^a_{\ b}+\Pi^a_{\ [d}\bnb_{c]}\bar{X}_b-\\
&-&\Pi_{b[d}\bnb_{c]}\bar{X}^a+\bar{Z}_{[d}\bnb_{c]}P^a_{\ b}+\bar{Z}_b\bnb_{[c}P^a_{\ d]}-\bar{Z}^a\bnb_{[c}P_{d]b}+
\bar{X}_{[d}\bnb_{c]}\Pi^a_{\ b}+\bar{X}_b\bnb_{[c}\Pi^a_{\ d]}-\\
&-&\bar{X}^a\bnb_{[c}\Pi_{d]b}
+\fr{1}{2}(P^a_{\ [c}\bar{Z}_{d]}\bar{Z}_b+\bar{Z}_r\bar{Z}^rP^a_{\ [c}P_{d]b}-\bar{Z}^aP_{b[c}\bar{Z}_{d]})+\\
&+&\fr{1}{2}(\Pi^a_{\ [c}\bar{X}_{d]}\bar{X}_b+\bar{X}_r\bar{X}^r\Pi^a_{\ [c}\Pi_{d]b}-\bar{X}^a\Pi_{b[c}\bar{X}_{d]}),
\enr
from which we get
\bnr
'L^0_{ab}&=&L^0_{ab}+(2-p)\bnb_a\bar{Z}_b-2\bar{Z}^r\bnb_rP_{ab}+\fr{1}{2}(p-2)(\bar{Z}_a\bar{Z}_b+\fr{1}{2}\bar{Z}_r\bar{Z}^rP_{bc})\\
'L^1_{ab}&=&L^1_{ab}+(2-n+p)\bnb_a\bar{X}_b-2\bar{X}^r\bnb_r\Pi_{ab}+\fr{1}{2}(n-p-2)(\bar{X}_a\bar{X}_b+
\fr{1}{2}\bar{X}_r\bar{X}^r\Pi_{bc}).
\enr
Plugging these expressions in the definitions of $'T^a_{\ bcd}$ and $T^a_{\ bcd}$ and using the above relations involving 
$'\bR^a_{\ bcd}$ and $\bR^a_{\ bcd}$ we find an equation similar to (\ref{cond-g}) but containing the difference $'T^a_{\ bcd}-T^a_{\ bcd}$ 
instead of the Lie derivative of $T^a_{\ bcd}$ and the variables $\bar{Z}_a$, $\bar{X}_a$ in place of $\bphi_a$, $\bchi_a$. From this 
point it is already easy to conclude 
$$
\ ^{||}\ 'C^a_{\ bcd}=\ ^{||}C^a_{\ bcd},
$$
as desired. Of course all this procedure can be repeated {\em mutatis mutandis} to get similar invariance rules for the tensors
$\bar{\La}^a_{\ bc}$, $\bnb_a\Pi^b_{\ c}$ and $\ ^{\perp}C^a_{\ bcd}$.
%
% BibTeX users please use
% \bibliographystyle{}
% \bibliography{}

\begin{thebibliography}{}
\bibitem{ANCIKOV} An\v cikov A. M. Izv. Vys\v s. U\v cebn. Zaved. Matematika \textbf{48}, (1965) 13-18.
(English translation: Amer. Math. Soc. Transl. \textbf{92}, (1970) 201-207.)  
\bibitem{BOMPIANI} Bompiani E. Rendiconti del Circolo Matematico di Palermo \textbf{48}, (1924) 121-134.
\bibitem{CAROT} Carot J. and Tupper B. O. J., Class. Quantum Grav. \textbf{19}, (2002) 4141-4166.
\bibitem{SERGI} Coll B., Hildebrandt S. R. and Senovilla J. M. M.
 Gen. Rel. Grav. \textbf{33}, (2001) 649-670.
\bibitem{EISENHART} Eisenhart L. P. \textit{Riemannian Geometry} (Princeton University Press, Princeton 1964) p. 33.
\bibitem{EISENHARTII} Eisenhart L. P. \textit{Continuous groups of transformations} (Dover publications Inc., 
New York 1933).
\bibitem{BI-CONFORMAL}
Garc\'{\i}a-Parrado A. and Senovilla J.M.M., Class. Quantum Grav. \textbf{21}, (2004) 
2153-2177.
\bibitem{HALL} Hall G. S., J. Math. Phys. \textbf{32}, (1991) 181-187.
\bibitem{LIBROHALL} Hall G. S. \textit{Symmetries and Curvature Structure in General Relativity} (World Scientific, Singapore 2004).
\bibitem{PETROV} Petrov A. Z. \textit{New methods in general relativity} (Nauka, Moscow 1966).
\bibitem{SCHOUTEN}
Schouten J. A. \textit{Ricci Calculus} (Springer, Berlin 1954).
\bibitem{YANO2} Yano K., \textit{Proc. Imp. Acad. Tokyo} \textbf{16}, (1940) 83-86.
\bibitem{YANO}
Yano K., \textit{Theory of Lie Derivatives} (North Holland, Amsterdam 1955). 
\bibitem{YANOMASAHIRO} Yano K. and Masahiro K. \textit{Structures on manifolds} (World Scientific, Singapore 1984).
\end{thebibliography}

\end{document}